\documentclass[twocolumn]{aastex63}
\pdfoutput=1 %for arxiv submission to use pdf
\usepackage{footnote}
\usepackage{float}
\usepackage[caption=false]{subfig}
\usepackage{gensymb}
\DeclareUnicodeCharacter{2212}{-}

\newcommand\blfootnote[1]{%
  \begingroup
  \renewcommand\thefootnote{}\footnote{#1}%
  \addtocounter{footnote}{-1}%
  \endgroup
}

\newcommand{\GHOSTcount}[0]{$16,175$ }

\begin{document}

\title{GHOST: Using Only Host Galaxy Information to Accurately Associate and Distinguish Supernovae}

\author{\href{https://orcid.org/0000-0003-4906-8447}{Alex Gagliano}}\blfootnote{Corresponding author: Alex Gagliano\\\href{mailto:gaglian2@illinois.edu}{gaglian2@illinois.edu}}
\affiliation{Department of Astronomy, University of Illinois at Urbana-Champaign, Urbana, IL 61801, USA}
\affiliation{National Center for Supercomputing Applications, Urbana, IL 61801, USA}
\affiliation{Center for Astrophysical Surveys, Urbana, IL, 61801, USA}
\affiliation{NSF Graduate Research Fellow}
%\collaboration{4}{(The DC superheros group)}

\author{Gautham Narayan}
\affiliation{Department of Astronomy, University of Illinois at Urbana-Champaign, Urbana, IL 61801, USA}
\affiliation{National Center for Supercomputing Applications, Urbana, IL 61801, USA}
\affiliation{Center for Astrophysical Surveys, Urbana, IL, 61801, USA}
%\collaboration{4}{(The DC superheros group)}

\author{Andrew Engel}
\affiliation{Department of Astronomy, University of Illinois at Urbana-Champaign, Urbana, IL 61801, USA}
%\collaboration{4}{(The DC superheros group)}

\author{Matias Carrasco Kind}
\affiliation{Department of Astronomy, University of Illinois at Urbana-Champaign, Urbana, IL 61801, USA}
\affiliation{National Center for Supercomputing Applications, Urbana, IL 61801, USA}
\affiliation{Center for Astrophysical Surveys, Urbana, IL, 61801, USA}
\collaboration{4}{(The LSST Dark Energy Science Collaboration)}

\begin{abstract}
We present \texttt{GHOST}, a database of \GHOSTcount spectroscopically classified supernovae and the properties of their host galaxies. We have constructed \texttt{GHOST} using a novel host galaxy association method that employs deep postage stamps of the field surrounding a transient. Our gradient ascent method achieves fewer misassociations for low-$z$ hosts and higher completeness for high-$z$ hosts than previous methods. Using dimensionality reduction, we identify the host galaxy properties that distinguish supernova classes. Our results suggest that the host galaxies of SLSNe, SNe Ia, and core-collapse supernovae can be separated by brightness and derived extendedness measures. Next, we train a random forest model to predict supernova class using only host galaxy information and the radial offset of the supernova. We can distinguish SNe Ia and core-collapse supernovae with $ \sim 70$\% accuracy without any photometric or spectroscopic data from the event itself. Vera C. Rubin Observatory will usher in a new era of transient population studies, demanding improved photometric tools for rapid identification and classification of transient events. By identifying the host features with high discriminatory power, we will maintain SN sample purities and continue to identify scientifically relevant events as data volumes increase. The \texttt{GHOST} database and our corresponding software for associating transients with host galaxies are both publicly available through the \texttt{astro\_ghost} package.
\keywords
{cosmology: observations, supernovae, machine learning, random forests}
\end{abstract}

\section{Introduction}\label{Intro}
In wide-fast-deep mode, Vera C. Rubin Observatory's \citep[Rubin Observatory;][]{2019Ivezic} Legacy Survey for Space and Time (LSST) will image the entire Southern sky every $\sim4$ nights. Much of this data will be rich in transient activity, with a predicted annual detection rate of 20,000 luminous ($M_v < −16$)  supernovae \citep{2009LSSTScienceBook}. Type Ia supernovae (SNe Ia) will be especially valuable discoveries, as their standardizability make them ideal for measuring cosmological distances and tracing the expansion history of the Universe. Since their initial use in discovering the accelerated expansion of the universe (\citealt{riess1998observational,Perlmutter1999}), SN Ia samples have placed strong constraints on both the dark energy equation of state (parameterized by $w$) and its potential evolution with redshift (\citealt{scolnic2018pantheon, 2018JonesSample}). 

LSST will discover supernovae photometrically, challenging our ability to maintain pure SNe Ia samples. Contamination from Ib and Ic core-collapse supernovae, whose light curves closely resemble those of SNe Ia, can systematically bias derived estimates for $w$ toward a time-evolving dark energy equation of state \citep{2010Contamination}. A select few supernovae discovered by LSST will be prioritized for rapid follow-up on smaller telescopes, where high-cadence spectroscopy can aid in classification and progenitor studies, but we will need accurate photometric classification to first identify this scientifically valuable subset.

To address this issue, a significant amount of effort has been devoted to developing accurate photometric classification algorithms. These classifiers primarily employ template fitting methods on real or simulated light curves (e.g. \citealt{poznanski2007bayesian, karpenka2013simple}), novel machine learning algorithms (\citealt{ishida2013kernel, kimura2017single, muthukrishna2019rapid, moller2020supernnova}), or a combination of the two \citep{lochner2016photometric}. Community-wide challenges have also drawn upon expertise from outside of the astronomical community, starting with the Supernova Photometric Classification Challenge (SNPhotCC; \citealt{kessler2010supernova}) in preparation for the Dark Energy Survey (DES) and closely followed by the Photometric LSST Astronomical Time-Series Classification Challenge (PLAsTiCC; \citealt{2019PLASTICC}) in preparation for LSST. Despite the accuracy of these previously developed methods, many classifiers are trained and tested on archival photometry spanning the full phase of previously discovered events; during LSST operations, however, the massive data volumes will require us to operate algorithms in real time and on incomplete light curves. This has led to the creation of brokers, automated systems that will rapidly ingest and process alerts from a variety of synoptic surveys such as LSST (e.g ANTARES, \citealt{narayan2018machine};  LASAIR, \citealt{smith2019lasair}). Marshals, platforms for automated coordination of telescope follow-up, are a valuable counterpart to these brokers. Marshals have existed for over a decade \citep{2009Law}, but their scope has expanded significantly in recent years to meet the growing discovery rate of events \citep{2019Kasliwal}. Marshals will soon operate in real-time and with automatic target selection (e.g. \citealt{2020Sravan}). Once brokers have reduced the stream of alerts and marshals have prioritized targets for observation, these objects are sent to Target and Observation Managers (TOMs; \citealt{2018Street}) for scheduling. This suite of software tools will be critical for streamlining communication between sites and reducing the latency for obtaining new observations.
% constructing multi-wavelength datasets for new events by
% With new pipelines for processing massive data streams, scientists will be able to rapidly identify and analyze new events as surveys come online. 

Real-time photometric classification will be necessary not only for maintaining pure SN Ia samples, but also for studying the progenitor physics of rare phenomena in detail (e.g. \citealt{2019De}; \citealt{2017Hosseinzadeh}; \citealt{2017Jiang}; \citealt{2020Miller}; among others). Several SNe Ia discovered at early times, including 2018oh (\citealt{Dimitriadis19:18oh_k2,Shappee2019a}), have revealed flux excesses in the first few days after explosion, providing a valuable probe of shock interactions with a companion star and with the circumstellar environment. Fast-evolving luminous transients (FELTs) represent an additional class of novel events with light curve rise times as short as 2.2 days (for KSN2015K; \citealt{rest2018fast}), and the number of fast transients discovered continues to be grow (\citealt{poznanski2010unusually, 2018Prentice, 2020Tampo}). Rubin Observatory will continue to push our explored parameter space to brighter cataclysmic events occurring across even shorter ($<1$ day) timescales. Rapid identification of the events within this parameter space will be critical for accurately characterizing their evolution and constraining progenitor models.  %Photometric classifiers that require the complete light curve of an event will be inadequate for understanding the progenitor physics of events within this parameter space; accurate real-time classification will be crucial.
In an effort to bridge this gap, a series of real-time classifiers are now being developed that leverage Recurrent Neural Networks (RNNs). These networks modify the weights from previous layers as data is added, allowing event classifications to update dynamically while targets are still being observed. These methods have achieved high accuracies distinguishing between supernovae using photometry pre-maximum (95\%, \citealt{muthukrishna2019rapid}; 83\%, \citealt{moller2020supernnova}), and RNNs trained directly on postage stamps have also shown promise for distinguishing a diverse range of transient and non-transient sources with limited phase coverage \citep{2019Carrasco}.
%\cite{muthukrishna2019rapid} implemented a deep recurrent neural network for event classification, achieving a mean accuracy of 95\% at early epochs in discriminating 12 transient classes before maximum.
%\citep{2019Carrasco} \textbf{TEXT HERE ABOUT OTHER CLASSIFIERS }
Despite these early successes, the current generation of real-time classifiers are trained on simulated data. This makes them sensitive to shot noise and additional artifacts in real photometric measurements at early times. Further, although LSST will image the entire sky every 4 nights in wide-fast-deep mode, each sweep will be done in a single passband. As a result, revisits \emph{using the same passband} are spaced between 10--20 nights in $grizy$, with even larger gaps in $u$. This presents a challenge for constraining the shape and color evolution of light curves of individual events, information upon which most classifiers depend.

By decreasing the dependence on early and sparse photometry, host galaxy information can help mitigate these issues. Supernova classes have been known to occur at different rates within different morphological classes of galaxies for decades, beginning with the association of type I supernovae with I0 galaxies \citep{oemler1979type}. More recently, statistical studies undertaken by \cite{foley2013classifying} and \cite{hakobyan2014supernovae} found that type Ia supernovae occur more frequently in early-type, red galaxies with low star formation rates than in late-type, blue galaxies with high star formation rates. In addition, \cite{kelly2012core} suggest that stripped-envelope SNe Ib/c are found mainly in metal-poor galaxies, and early-type galaxies hosting SNe II/Ib are on average bluer than early-type galaxies hosting SNe Ia \citep{suh2011early}. Correlations have also been identified between supernova type and both host galaxy morphology \citep{foley2013classifying} and global star formation rate (\citealt{kelly2014hosts, zhou2019local}).

Supernova classifiers can directly benefit from these host galaxy correlations. \cite{foley2013classifying} found that a Naive Bayes classifier using \emph{only} host galaxy morphology, absolute magnitude, color, galactic offset and pixel rank could classify supernovae as accurately as the best light curve classification methods of that time. More recently, \cite{baldeschi2020star} showed that photometric estimates of host galaxy star formation can be used to increase the purity of SN Ia and core-collapse supernova samples by 10\% and 20\% from random guessing, respectively. We extend these previous efforts by considering an exhaustive list of photometric host galaxy properties and a sample of host galaxies greater than an order of magnitude larger than previous studies. The single-visit depth of LSST ($r\sim24.5$ mag) will allow us to identify transient host galaxies out to redshift $\sim$1, and we can use the photometry of these galaxies as \textit{a priori} data to improve early-time classification.

A comprehensive understanding of the dominant correlations linking a transient to its host galaxy will improve our ability to reduce systematic uncertainties within SN Ia data. SN Ia explosions are known to be influenced by their host environments, as metallicity \citep{hoflich2010secondary} and extinction from dust \citep{mandel2017type} can change the shape and peak magnitude of the resultant light curve, respectively. Moreover, many groups have identified correlations between SN Ia Hubble Residuals and global properties of the host galaxy. These correlations include morphology \citep{Hamuy1996}, metallicity \citep{d2011spectroscopic}, and star formation rate derived from ground-based SDSS and Gemini observations \citep{henne2017influence}. Only weak correlations have been found to date; if stronger correlations exist, they could be identified through a statistical study of host galaxy properties spanning both the global and the local scales. Unfortunately, the paucity of higher resolution imaging has limited SN host galaxy studies at the pixel level and at high redshift. Our limited understanding of these correlations has resulted in an \textit{ad hoc} piecewise ``Hubble-mass-step" correction to Hubble Residuals (\citealt{kelly2010hubble, sullivan2010dependence, 2010Lampeitl, kim2014type}). The SN Ia mass-luminosity correlation will remain even as other systematic effects are addressed by Rubin Observatory's all-sky photometric calibration, and it will soon become a dominant source of uncertainty \citep{scolnic2018pantheon}. Characterizing these correlations now will be essential for strengthening our ability to probe dark energy and the expansion rate of the universe.

In this work, we explore the use of host galaxy information in classifying supernovae. We extend the work done by \cite{foley2013classifying} by constructing a database of host galaxy-supernova pairs greater than an order of magnitude larger than the Lick Observatory Supernova Search (LOSS; \citealt{2000Li}) sample they use. We also consider $>140$ additional host galaxy features. In addition to our data products, we have developed a new method for identifying transient host galaxies in deep surveys using postage stamps of the field. This method achieves accuracies superior to the commonly used directional light radius method at low-$z$, and with greater completeness at high-$z$ where little morphological information is available. Our dataset for this study contains the majority of previous spectroscopically classified supernovae and their host galaxies, and this sample is intrinsically biased toward low-redshift events (which are more easily detected). Because LSST will expand our discovery space to higher redshift, our database will need to be augmented with newly discovered SNe for \texttt{GHOST} to provide accurate classifications on the LSST alert stream. Our pipeline for host galaxy association will soon be integrated into the alert stream of the Young Supernova Experiment (YSE), which observes a total survey area of 1500 square degrees in two bands per epoch to discover and study supernovae before peak brightness (\citealt{2019JonesYSE}; \citealt{2020JonesYSE}).  

\begin{figure}
    \centering
    \includegraphics[width=1.1\linewidth]{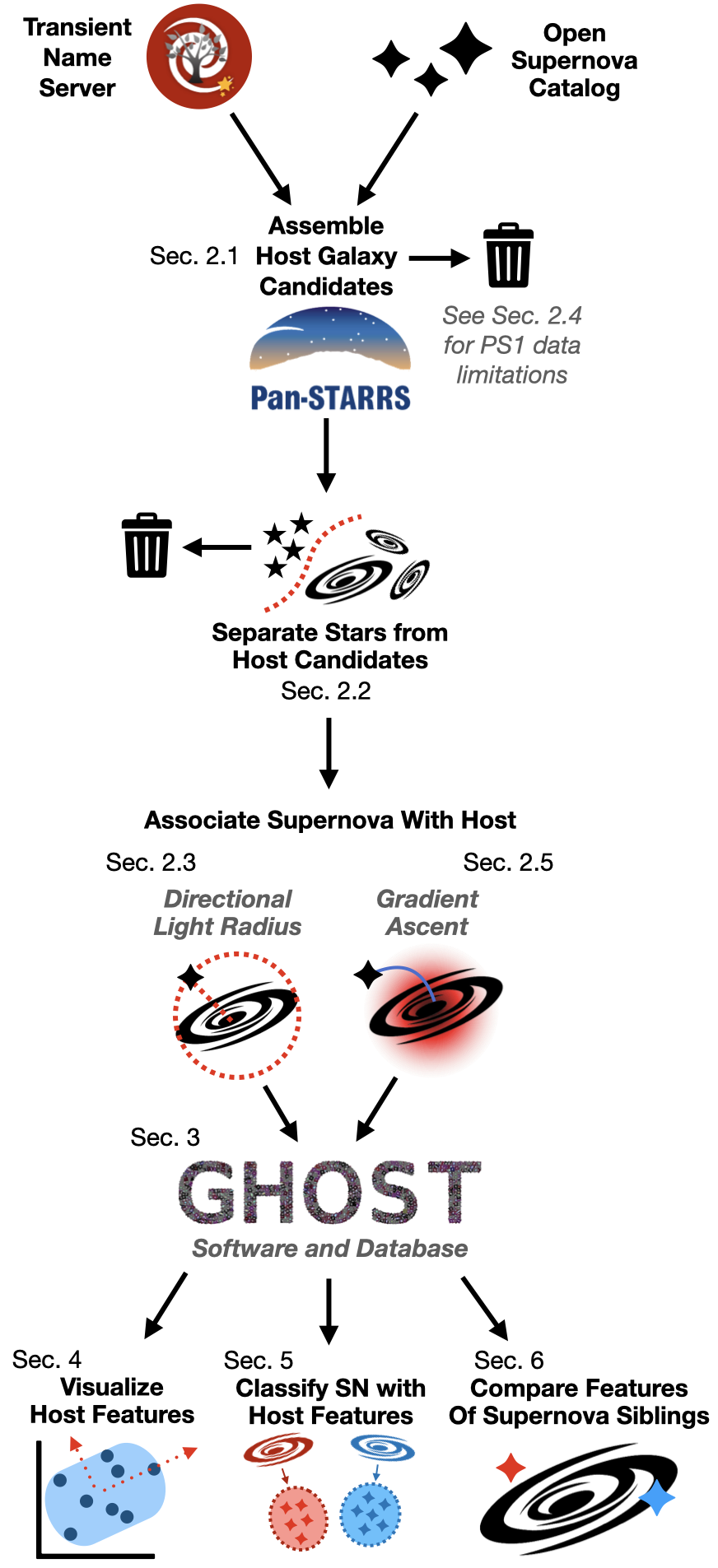}
    \caption{An outline of the analysis completed in this work, with corresponding sections labeled. The primary data sources for our supernova sample are given at the top of the flowchart, and major steps within our pipeline are indicated with text.}
    \label{fig:GHOSTflowchart}
\end{figure}
%\newpage
\subsection{Structure of this Paper}\label{Outline}

We provide a schematic overview of this work in Figure \ref{fig:GHOSTflowchart}. Our paper is laid out as follows: In \textsection \ref{Methods}, we describe our methodology for constructing and validating our database of supernovae and their host galaxies. In \textsection \ref{softwareSection} we outline the data and software products that accompany this work. These products can be used to query the released database and identify the host galaxies of new transients. We compare PS1-derived brightness and morphology features of host galaxies using Principal Component Analysis in \textsection \ref{DimensionReduction}.  We then compare the host galaxies of underrepresented supernova classes using t-Distributed Stochastic Neighbor Embedding (tSNE) in \textsection \ref{tSNEResults}. Next, we introduce our algorithm to classify supernovae and present results in \textsection \ref{ClassificationMainSection}. Section \textsection \ref{siblings} is devoted to the analysis of supernovae originating in the same host galaxy. We conclude by summarizing our results in \textsection \ref{Discussion} and discussing future directions for the research in \textsection \ref{FutureWork}.

\section{Host Galaxy Identification}\label{Methods}
To construct a dataset of supernova host galaxies, we first assemble a dataset of spectroscopic supernovae. We have downloaded all spectroscopically classified supernovae from the Transient Name Server (TNS; \url{https://wis-tns.weizmann.ac.il/}) and the Open Supernova Catalog (OSC; \url{https://sne.space/}). After removing events with identical names and positions, we are left with 20,736 events. We then implement a series of quality cuts and a novel matching algorithm to identify the host galaxies of supernovae in our sample. Our final database contains host galaxy information for \GHOSTcount supernovae -- just over 78\% of all spectroscopically classified events at the time of writing. We outline our methodology for matching supernovae with their host galaxies below.
    
\subsection{PS1 Querying for Host Candidates}\label{Methods:PS1Querying}
We use the first data release of Pan-STARRS (PS1; \citealt{chambers2016pan}) to search for candidate host galaxies due to its survey depth (5$\sigma <$ 23.3 in $g$-band) and extensive list of sources ($>$3B unique objects). We have chosen Pan-STARRS data over SDSS for its superior sky coverage and comparable depth (SDSS has a median 5$\sigma$ depth of $23.13$ in $g$), and over alternative sky surveys for its resolution ($0.25\arcsec$/px, two orders of magnitude higher than the 21$\arcsec$/px resolution of the Transiting Exoplanet Survey Satellite). We opt for the first data release to take advantage of the \textit{bestDetection} flag, which was corrupted in DR2. 
    
To first order, a host galaxy can be found by cross-matching the redshift of the event with the redshifts of nearby sources; however, the majority of supernovae discovered prior to 2010 do not list a spectroscopic redshift in TNS, and the NASA Extragalactic Database (NED\footnote{\url{https://ned.ipac.caltech.edu/}}) is only 75\% complete for redshifts of galaxies at $z \leq 0.05$. The catalog is even less complete at higher redshifts \citep{kulkarni2018redshift}. For these reasons, supernovae and their host galaxies cannot be associated by redshift alone. Instead, we construct a table of potential host galaxies for each transient event using the following procedure:

    \begin{enumerate}
        \item Query Pan-STARRS for all catalogued objects within $30\arcsec$ of the supernova. If a host galaxy has been reported in TNS or OSC, we instead take its coordinates as the center of our cone search. The resultant table is likely to contain artifacts and other objects that are irrelevant to this analysis, such as HII regions. 
        \item Use the PS1 \textit{primaryDetection = 1} flag to remove duplicate detections of the same source. This is common in fields containing a single large galaxy with a spatially resolved core or many associated HII regions. We identified multiple faint galaxies without a \textit{primaryDetection = 1} entry at this stage, and so we caution that this cut may preferentially associate supernovae with nearby large galaxies. 
        \item Remove any object not detected in $gri$. We have found that the number of artifacts in our candidate host galaxy list increases dramatically without this cut, but we do not cut on $z$ or $y$ bands so that we retain high-$z$ host galaxies in our sample. 
        \item Remove sources with \textit{bestDetection = 0} if at least one source with \textit{bestDetection = 1} is present in the field. If no best detection sources exist, do not eliminate any potential host galaxies for that supernova. We had initially removed all sources with \textit{bestDetection = 0}, but found that this removed a non-negligible fraction of plausible host galaxies.
        \item Remove sources with \textit{qualityFlag = 128}  (indicating a poor-quality stack object).
    \end{enumerate} 
    
    We list the fraction of supernovae removed at each of these steps in Table \ref{tab:Cuts}. After removing the majority of artifacts from our table, we eliminate the PS1 sources corresponding to stars.
    \subsection{Star/Galaxy Separation for Deep Surveys}\label{Methods:StarRemoval}

      \begin{figure*}[!ht]
     \centering
\includegraphics[width=\linewidth]{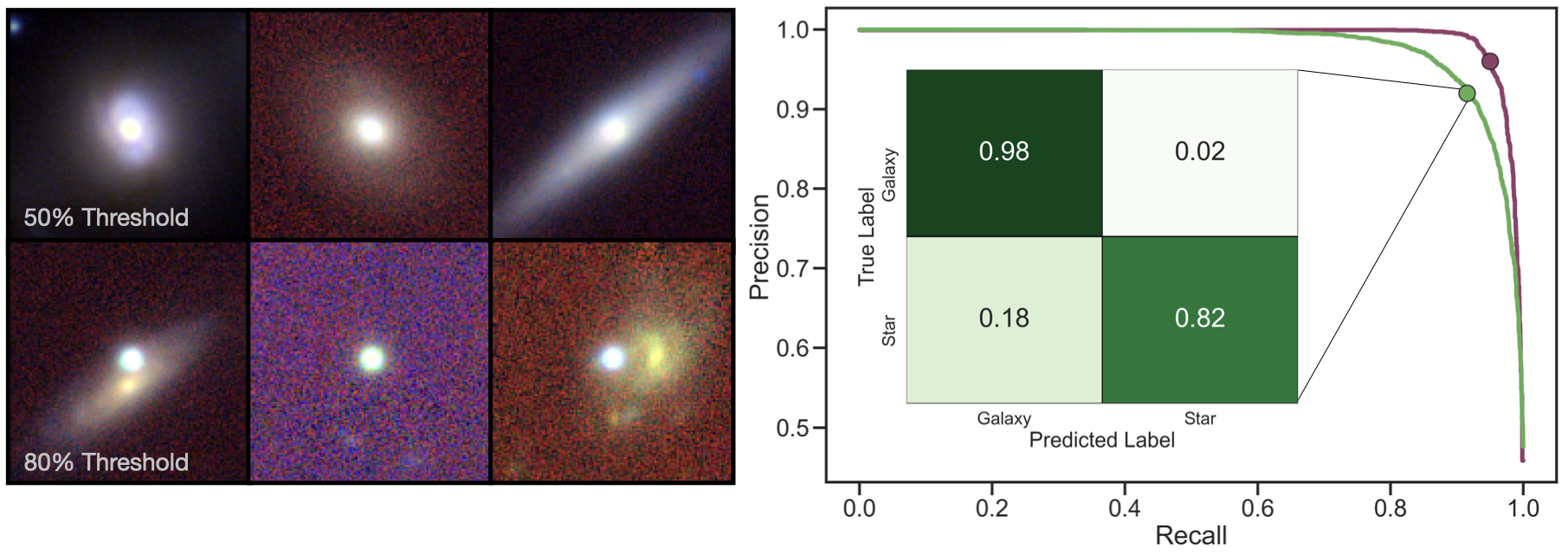}\label{fig:StarsVGals}
     %\vspace*{-5mm}
     \caption{ \textbf{Left.} Postage stamps for a few RF-identified stars in our dataset using a decision threshold of 50\% (above) and 80\% (below). Although the 50\% model achieves high accuracies across the full dataset, it is more likely to erroneously identify bright galactic nuclei as stars. By increasing the decision threshold to 80\%, the RF model becomes less likely to flag these sources. \textbf{Right.} The precision-recall curves for our RF model. The points top right indicate our final decision threshold. The green (purple) line corresponds to the full (masked) sample, where the masked sample contains only sources with low photometric uncertainty ($MagErr < 0.02$) and greater than 3 detections in $griz$. With a star/galaxy decision threshold of 80\%, we achieve a recall and precision of 92\% for the full sample and 95\% and 96\%, respectively, for the masked sample. The inset confusion matrix presents our final classification results on the full sample. Because we have biased our algorithm against removing potential hosts, we achieve a 2\% false negative rate for our galaxy classification with negligible impact on the mean accuracy, precision, and recall of the model.} \label{fig: Gals_v_Stars_Histograms}
    \end{figure*}
    A useful feature for distinguishing point-like and extended sources is brightness measured in different apertures \citep{slater2020morphological}. The total flux of an unresolved source can be captured equally well by a point-spread function (PSF) model or by a more complicated aperture model such as Kron, which defines the radius of its aperture as $2.5 \times$ the first radial moment of a source \citep{magnier2016pan}. To first order, stars can be separated from galaxies with a horizontal cut in $m_{\rm PSF} - m_{\rm Kron}$ space (See section 6.3 of \citealt{chambers2016pan}), where stars will cluster around $m_{\rm PSF} - m_{\rm Kron} = 0$. This cut introduces a bias against distant high-redshift galaxies, which are well-mixed with stars in this space faintward of $m_{\rm PSF}\sim$21. Further, stars brighter than  $m_{\rm PSF}\sim$15 will shift above this line because of saturation, and a horizontal cut will not capture these bright sources.
      
      \begin{figure}
          \centering
          \includegraphics[width=\linewidth]{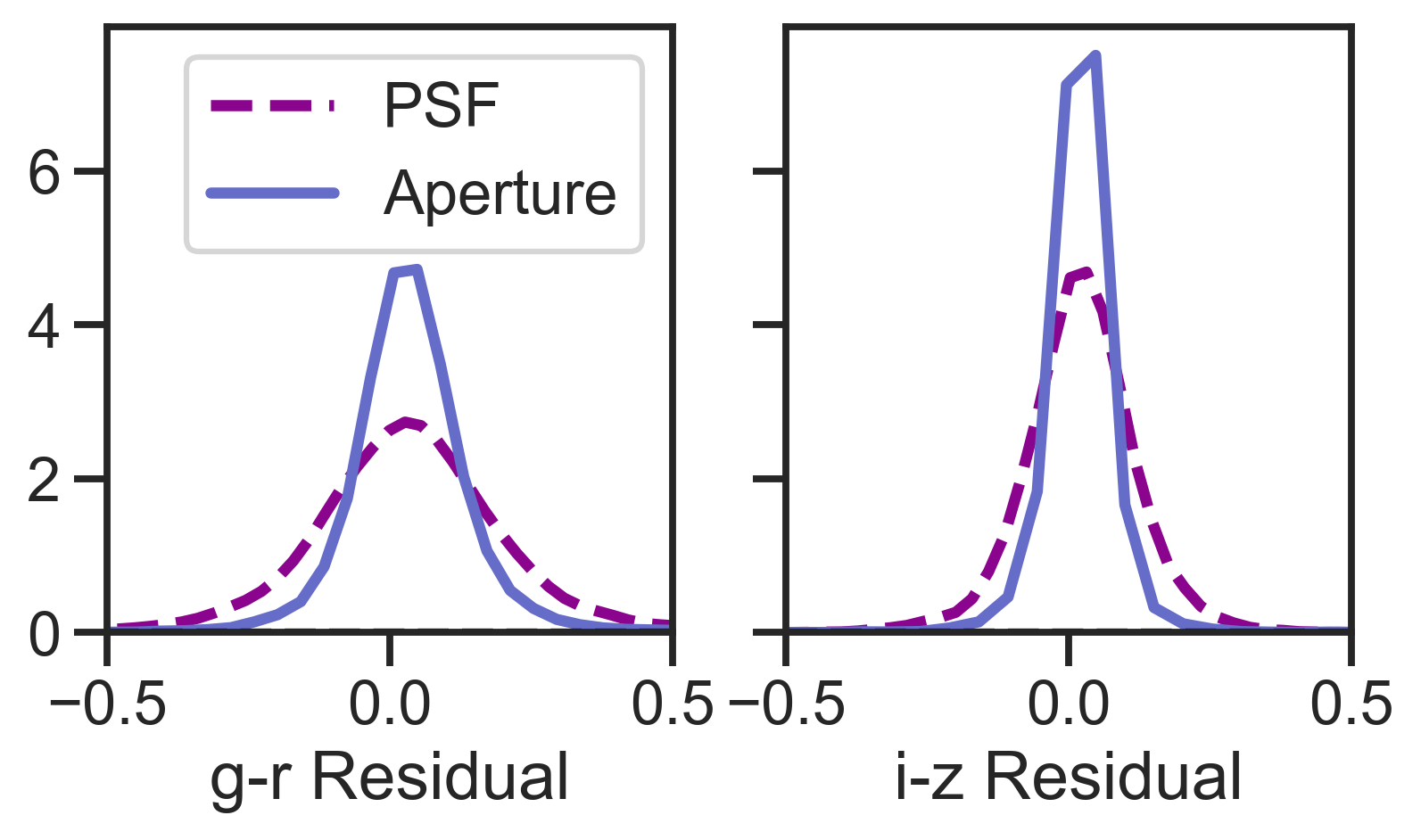}
          \caption{The residuals in $g-r$ and $i-z$ colors calculated between the PS1 stellar locus and identified stars within our candidate host galaxy table. Colors have been corrected for extinction using the reddening values and extinction coefficients from \cite{schlafly2011measuring}.}
          \label{fig:ColorResiduals}
      \end{figure}
      To more accurately distinguish stars from potential host galaxies, we query the PS1-Source Types and Redshifts with Machine learning (PS1-STRM) catalog from \cite{2020MNRAS_STRM} for every source within our candidate list. The PS1-STRM catalog uses a neural network to label sources as stars, galaxies, and QSOs. Out of $\sim$49,500 unique sources, nearly 48,000 belong to one of these categories, whereas the remaining 1,500 are listed as `unknown'. To classify this remaining subset, and to avoid querying this database each time our pipeline is run, we implement a Random Forest (RF; \citealt{breiman2001random}) algorithm to separate stars from galaxies and QSOs using features of PS1-STRM-classified sources. A Random Forest is a supervised learning algorithm composed of an ensemble of decision trees. In the classification problem, a random subset of input features is used to construct each decision tree and predict the class of a sample. Once the trees are constructed, the algorithm assigns a final class with a majority vote of the output from each tree. We use the random forest implementation in \texttt{sklearn}. For our final model, we require at least 8 observations to split a branch, two features per split, and a minimum of two observations per leaf (the end of each tree).
      %This feature bagging allows us to estimate the importance of features by their discriminatory power across many trees.
      %a maximum tree depth of 80,
      We train our RF model on the aperture magnitude ($m_{\rm Ap}$) of our potential host galaxies in $g$, $r$, and $i$; $m_{\rm Ap} - m_{\rm Kron}$ in the same bands; and the 4-dimensional color distance in $g-r$, $r-i$, $i-z$, and $z-y$ from the PS1 stellar locus, the path traced by stars in color-color space. Because of its generality, the stellar locus is a valuable tool for photometric calibration of large surveys \citep{high2009stellar}. We can also use it to help distinguish stars and galaxies, since stars within our sample should closely trace the locus whereas galaxies will not. Aperture magnitude is determined by integrating over the analytic PSF aperture, then extrapolating the PSF model to estimate the flux that was missed. We have chosen this measurement in place of $m_{\rm PSF}$ because it accounts for local variations in the PSF, and as a result its values have less dispersion (see Fig. \ref{fig:ColorResiduals}). We adopt a cubic spline with knots given by \cite{tonry2012pan} and an $r-i$ bin width of $0.001$ as our PS1 stellar locus. To calculate the 4-dimensional distance of each source from this locus, we use the equation \citep{covey2007stellar}:\begin{equation}
          4DCD = min_{k} \sum_{i=0}^4 \frac{\left(C_{i} - t_{i,k}\right)^2}{\sigma_{i}^2}
      \end{equation}
     where $C_i$ represents the $i$th color of a source, $t_{i,k}$ represents each value along the stellar locus for that color, and $\sigma_{i}$ represents the PS1 uncertainty in the source's color (calculated by combining the magnitude uncertainties in quadrature). The $4DCD$ is the minimum value of this equation calculated between all $k$ points along the stellar locus. This value approaches the orthogonal distance between a source and the stellar locus in 4D space as the number of discretized locus points approaches infinity.
     
     We considered the use of Gaia DR2 parallaxes as an additional feature within our RF model, as it would allow us to distinguish between galactic stars and distant galaxies; however, the depth of the survey (with a limiting magnitude of $G$ = 21) limits its completeness for a significant number of our sources. Further, the parallaxes reported in DR2 have significant uncertainties, and training our RF to identify stars by their large parallaxes would remove a non-negligible fraction of galaxies from our sample as well.
     
     To quantify the performance of our star-galaxy separator, we divide our STRM-classified data into 70\% training and 30\% testing sets, respectively. Next, we re-balance the training set using the Python package \texttt{Imbalanced-learn}. This prevents our algorithm from improving accuracy by classifying every source as the most populous class in the training data. Our training set consists of 12,560 stars and 12,560 galaxies/QSOs after re-balancing, and our testing set consists of 5,345 stars and 6,351 galaxies/QSOs. This brings our train and test fractions to 68.2\% and 31.8\%, respectively.
     
     We now consider the accuracy, precision, and recall of our model. The accuracy describes the total fraction of the test sample that is correctly labeled, and for a given class the precision is defined as
     \begin{equation}
        p = \frac{TP}{TP + FP}
     \end{equation}
     where the true positive rate $TP$ describes the number of objects correctly identified in this class and the false positive rate $FP$ describes the number of objects incorrectly labeled as members of this class. Recall is defined as 
     \begin{equation}
         r = \frac{TP}{TP + FN}
     \end{equation}
     where the false negative rate $FN$ denotes the number of objects incorrectly labeled as \textit{not} belonging to this class.
     
     Our model achieves a mean precision, mean recall, and mean accuracy of 93\% each distinguishing stars and galaxies/QSOs from this test set. We also create an ``ideal" sample of stars and galaxies by removing from our dataset any sources with missing brightness information, $ApMagErr > 0.02$, $ApMag > 20$, or fewer than 3 detections in $griz$. For the STRM-classified stars, we also remove sources with $psfQfPerfect < 0.9$, which would indicate that a non-negligible fraction of pixels were masked in the stack detection of this source. After splitting into training and testing samples and re-balancing the training set as before, this curated sample consists of 2,569 galaxies/QSOs and 2,569 stars in the training set, and 2,559 galaxies/QSOs and 1,101 stars in the testing set. Using the same model as before, our mean precision, mean recall, and mean accuracy each increase to 96\%. 
     %We find false-positive rates of 6.8\% and 3.6\% for the full and ideal samples, respectively, and false-negative rates of 8.0\% and 2.7\%.
     
     While this model achieves high accuracies on both our full and curated samples, the precision and recall of stars and galaxies should not be weighted equally. Retaining host galaxy candidates is significantly more important than removing all stars, since remaining stars may be removed by later cuts but it is highly unlikely that a galaxy will be added back to the data once it is removed. For this reason, we only remove sources that are classified as stars to $>$80\% probability by our RF model (instead of the fiducial 50\% cut). With this threshold, our mean accuracy on the full dataset decreases to 92\% (a decrease of 2\%), but our precision for stars rises to 98\% and our recall for galaxies reaches 99\%. Our mean precision on the curated sample is now 95\% (a decrease of 1\%) but our mean precision is unchanged. By minimizing the number of galaxies misidentified as stars, we achieve a classifier that is both accurate and conservative. We show the precision-recall curve for our classifier in Fig. \ref{fig: Gals_v_Stars_Histograms}, along with the final confusion matrix after increasing our decision threshold.
     
     We then remove the STRM-identified stars and the RF-predicted stars from our table of potential host galaxies. Galaxies with the smallest separation from the stellar locus and the stars with the largest locus separation were inspected using PS1 postage stamps, but no manual re-association was done at this stage. The RF-identified galaxies with smallest separation to the locus were nearly point sources, but were kept in the dataset due to the likelihood that they were QSOs.

     \subsection{The Directional Light Radius for Host Galaxy Association}\label{Methods:DLR}
      Supernovae are often associated with a host galaxy using the directional light radius (DLR) prescription outlined in \cite{gupta2016host}, where the angular separation $\theta$ is scaled by the radius of the host galaxy in the direction of the event $d_{DLR}$. A supernova embedded in a large galaxy is more likely to be found further from its host galaxy's nucleus than a small galaxy, and this normalized distance permits a direct comparison of supernova separations between host galaxies with vastly different scales. Normalized separation metrics have already been adopted by the Supernova Legacy Survey (SNLS; \citealt{sullivan2006rates}) and the SDSS-II Supernova Survey (SDSS-SNS; \citealt{2018SDSS}). As the rate of supernova detections continues to increase, the success of the DLR method at low-$z$ and in crowded fields has made it the preferred method for automated host galaxy association (see \citealt{gupta2016host} for a detailed review). 
      
\begin{figure}
    \centering \hspace*{-15mm}
    \includegraphics[width=\linewidth]{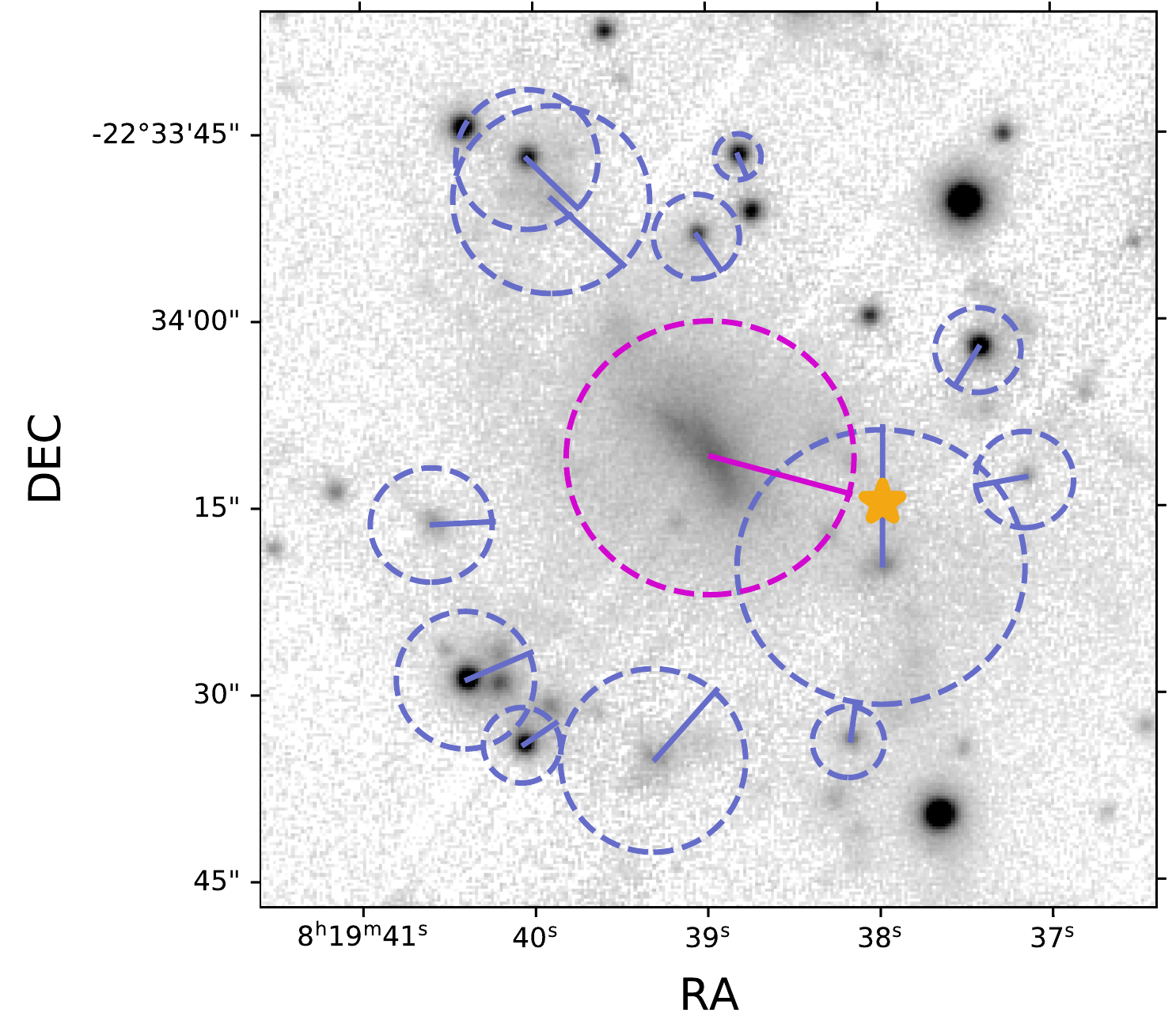}
    \caption{The PS1 postage stamp for SN 2014dp, with host galaxy candidates circled. We show the radius of each candidate in the direction of the transient, calculated using the Kron radius of the host galaxy and the DLR method in \textsection\ref{Methods:DLR}. Despite the center of the true host galaxy lying further from the transient event than other spurious candidates, the true host galaxy (magenta) is correctly identified using our modified DLR method. The match is verified using redshift information from NED.}%
    \label{fig:DLRexample}%
    \end{figure}

      The DLR method proceeds as follows. For a supernova located at $(x_{SN}, y_{SN})$ and a potential host galaxy located at $(x_{Gal}, y_{Gal})$, the Stokes parameters $U$ and $Q$ of the galaxy are given by its flux-weighted second order moments in PS1:
      \begin{equation}  
          U = M_{XY} ; \ \ \  Q = M_{XX} - M_{YY} 
      \end{equation}
      The angular tilt $\phi$ of the galaxy relative to celestial north is then found with
      \begin{equation} 
          \phi = \frac{1}{2} tan^{-1}(U/Q)
      \end{equation}
      Next, we calculate the aspect ratio of the host galaxy with 
      \begin{equation} 
            r_{a/b} =  \frac{(1 + \kappa + 2 \sqrt{\kappa})}{1 - \kappa} 
      \end{equation}
      where $\kappa$ is derived from the two Stokes parameters: \begin{equation} \kappa = Q^2 + U^2 \end{equation} 
      The angle $\gamma$, describing the angle of the supernova position relative to the galactic center and celestial north, is then found with
      \begin{equation}
           \gamma = tan^{-1}\frac{y_{SN} - y_{Gal}}{x_{SN} - x_{Gal}}
    \end{equation}
    Finally, we arrive at the angle $\beta$ subtended by the galactic semi-major axis and the vector connecting the supernova to the galactic center:
   $$ \beta = \phi - \gamma
    $$
    The DLR is found with these parameters and the galaxy's semi-major axis $r_{a}$ using the equation 
    \begin{equation}
     d_{DLR} = \frac{r_a}{\sqrt{(r_{a/b} \textrm{sin}\beta)^2 + (\textrm{cos}\beta)^2}}
     \end{equation}
     The scaled directional light radius is found by $\theta/d_{DLR}$, where $\theta$ is the Great Circle distance of a supernova from the center of its host galaxy in arcseconds. An illustration of this method is given in Figure \ref{fig:DLRexample}. 
     
    The DLR method uses second-order moments for estimating $U$ and $Q$, which are model independent; however, estimating the host galaxy's semi-major axis $r_a$ may require us to adopt a light profile model. We have selected the Kron radius, defined as $2.5 \times$ the first radial moment of a candidate host galaxy's surface brightness profile, in the band which has the highest SNR, as our value for $r_a$ to minimize this dependence. The mean $r$-band Kron radius for all potential host galaxies at this stage is $7.5\arcsec$. 
    
    \cite{gupta2013understanding} estimates that 7\% of matches in his sample found using this method are erroneous, with an additional 3\% of SNe left unassociated due to unreported second order moments. Using a naive DLR association, we find our fraction to be significantly higher. This is due to a combination of non-galaxy sources in our database and multiple PS1 entries for a single source, both of which were only partially removed in previous steps. To mitigate this issue, we have modified the DLR method outlined above. For each supernova in our sample, we complete the following steps:
    \begin{enumerate}
        \item Find the Kron radius of each potential host galaxy in the band with highest SNR. Then, calculate $d_{DLR}$. If either the Kron radius or the second-order moments are not provided, remove this source from the list of potential host galaxies. Repeat for all potential host galaxies.
        \item Eliminate host galaxies with $\theta/d_{DLR} > 5$. If no host galaxies remain after this step, the supernova is reported to be hostless and added to the sample of hostless events. 
        \item Rank-order the host galaxies by ascending values of $\theta/d_{DLR}$.
        \item If the remaining list contains at least one NED-identified galaxy, select the galaxy with the lowest $\theta/d_{DLR}$ as host. If not, select the source with the lowest $\theta/d_{DLR}$.
    \end{enumerate}
    
    \subsection{Limitations in the DLR Methodology Applied to PS1 Data}\label{Methods:PS1Issues}
    The depth of PS1 has allowed us to construct a complete list of supernova host galaxies, but several issues can arise in a survey as complete as PS1. First, de-blending errors plague many of the low-$z$ host galaxies in our sample. These galaxies are often characterized by several overlapping PS1 
    ``sources", and neither the \textit{primaryDetection} nor the \textit{bestDetection} flags are able to unambiguously identify the PS1 entry closest to the true galaxy center.
    
    Second, deep PS1 imaging is able to resolve dim host galaxy features that would be missed in other surveys. This introduces additional ``sources" to the list of candidate host galaxies for a supernova, further crowding low-redshift fields.  Because of the de-blending errors associated with highly extended sources, we also find non-physical Kron radius measurements for the majority of these sources. We often find a PS1 source near the true galactic center with an uncharacteristically low Kron radius, if one is reported at all. Conversely, sub-structures such as HII regions and bright stars may ``adopt" the light profile of an entire galaxy and report uncharacteristically high Kron radius values. These issues bias the DLR algorithm \textit{away} from large host galaxies where visual association would be trivial. 
    
    Third, high-redshift host galaxies can easily be mistaken for imaging artifacts, and are preferentially removed by the \textit{bestDetection} flag. The Kron radius for these objects is often (understandably) not reported. This leads to an overabundance of unassociated supernovae in sparse fields.
    
    We have attempted to develop a pipeline that is fault-tolerant to these limitations in PS1 data, but accurate radius estimates are critical to the success of the DLR method in crowded fields. \cite{gupta2013understanding} uses an upper limit of $\theta/d_{DLR} = 4$ as a balance between sample purity and efficiency. We have extended this value to $5$ to account for the low Kron radius estimates associated with low-redshift host galaxies; we note that a similar threshold is used for associating host galaxies in SNLS \citep{sullivan2006rates}.  
    
     By visually inspecting a subset of associated host galaxies, we predict our overall misassociation rate with the full pipeline to be $\sim5\%$; however, for low-redshift objects this rate can be as high as $30\%$ for the reasons given above. When DLR fails to find the true host galaxy, a supernova is often matched to an HII region or bright star. To maintain the accuracy of our associations at low redshift, we have developed a novel host galaxy association algorithm using the light profiles of sources near the supernova of interest. We provide a detailed description of this method below. 
 \subsection{Host Galaxy Association using Background Image Gradients}

\begin{figure*}[!ht]
  %  \captionsetup[subfigure]{labelformat=empty}
     \centering
     \subfloat[][]{\includegraphics[width=0.5\linewidth]{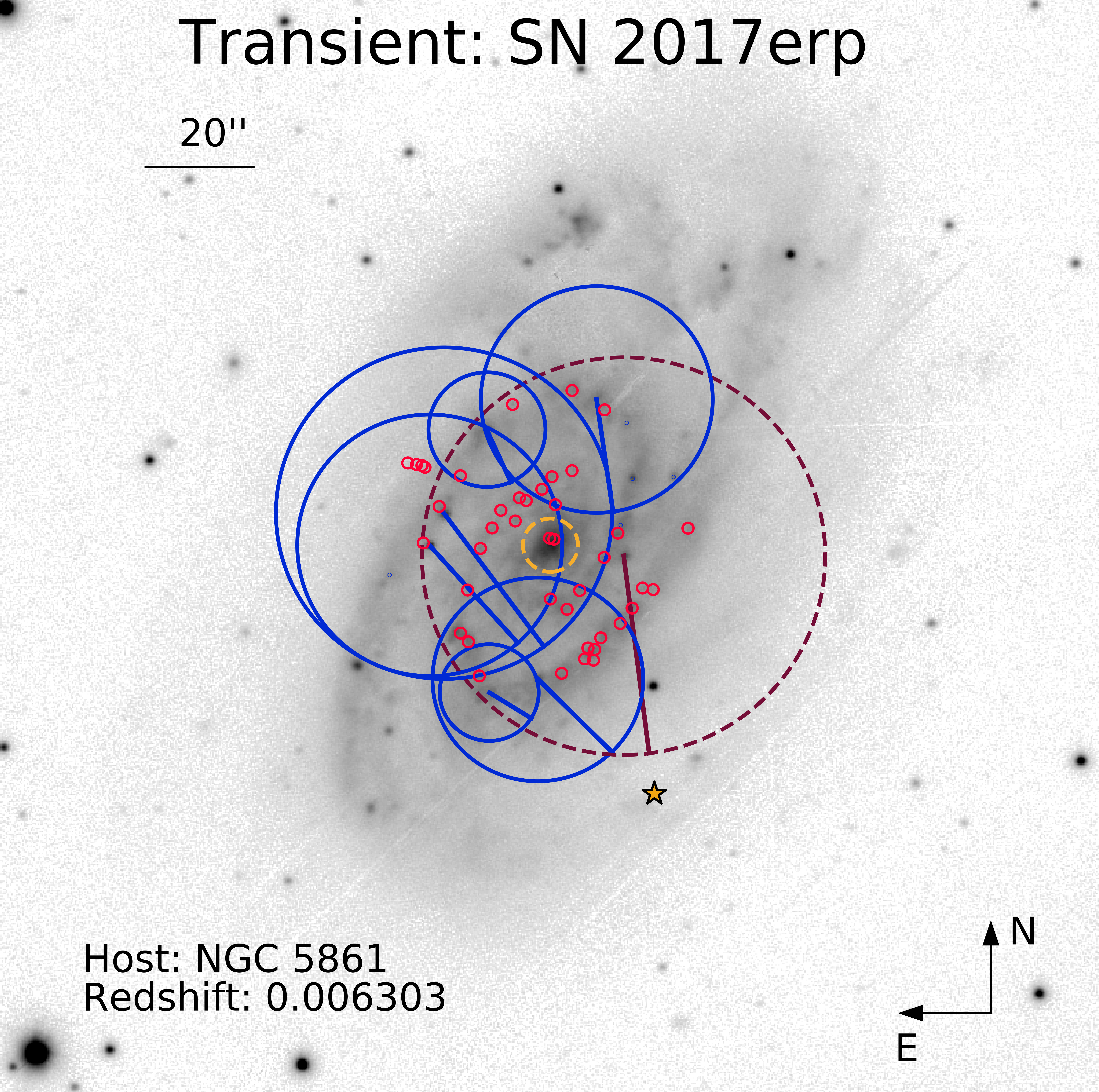}\label{fig:background_SN2017erp}}
     \subfloat[][]{\includegraphics[width=0.51\linewidth]{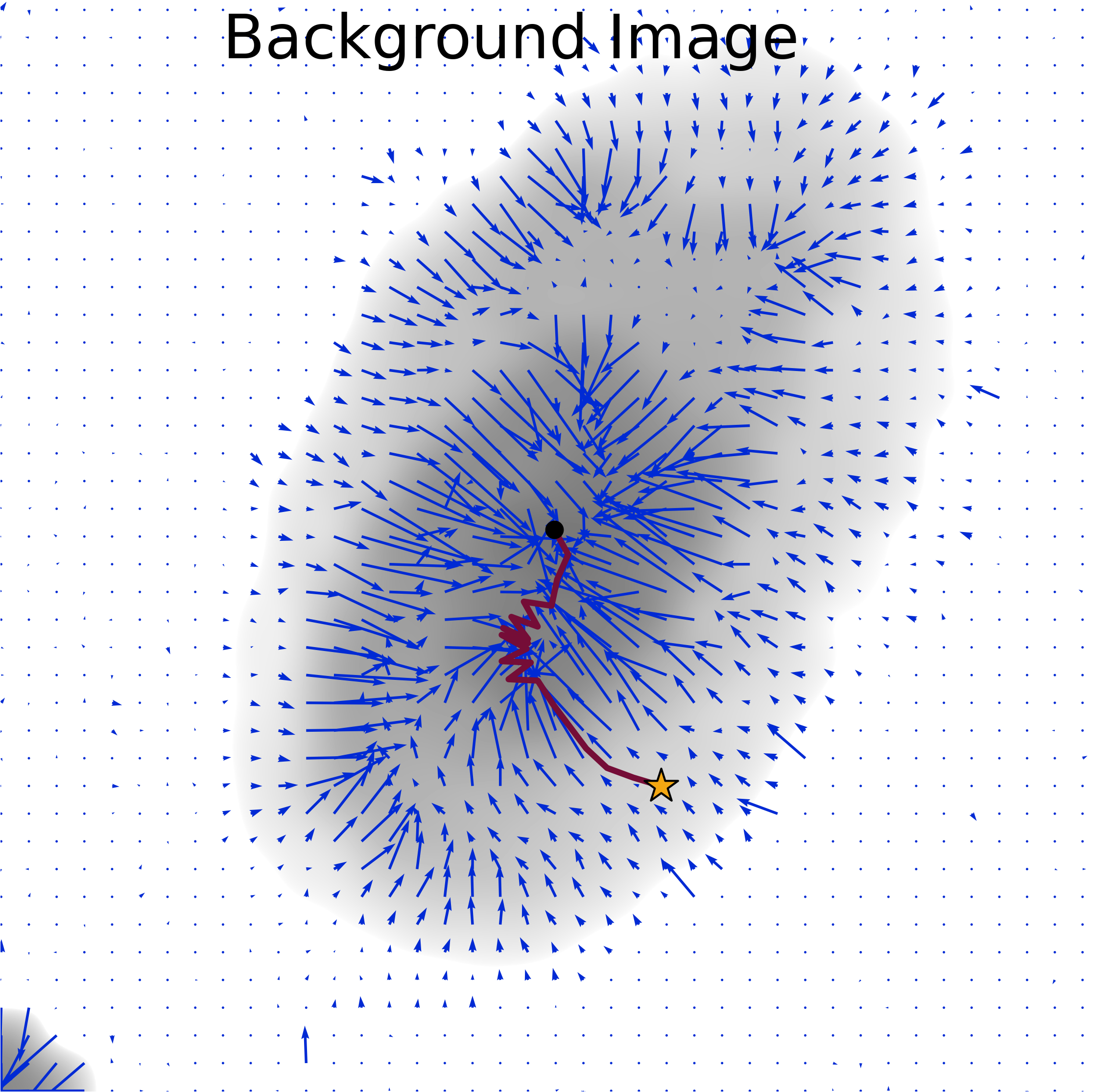}\label{fig:quiverDLR_SN2017erp}} 
     %\vspace*{-5mm}
     \caption{\textbf{a.} The gradient ascent algorithm applied to SN2017erp. The red circles correspond to the positions of PS1 sources eliminated before the DLR calculation, while the blue circles (scaled by $d_{DLR}$) denote potential host galaxies at the DLR stage. The dark magenta dashed circle denotes the DLR-selected host galaxy, while the smaller dashed gold circle at center denotes the PS1 source selected by gradient ascent. The position of the supernova is given by the yellow star. Despite numerous spurious detections in the field, the gradient ascent algorithm is able to rapidly identify the true host galaxy, which was later verified by redshift. \textbf{b.} The estimated background light profile of the central host galaxy, found by dividing the postage stamp into squares and applying a median filter to each of these sub-regions. We give the background pixel gradients, in blue, at every 20th pixel for clarity. The path traversed by the gradient ascent method is marked in dark magenta, and the black circle marks the final location reached. By masking saturated pixels and point-sources, we recover the light profile from the true host galaxy.}
     \label{fig:FullROC}
\end{figure*}
While deep imaging of extended host galaxies in PS1 makes source de-blending difficult, it makes visual association of SN host galaxies easier. We can rapidly identify a host galaxy by eye by first determining if the supernova is embedded within a galaxy's light profile and, if so, by tracing this profile from the location of the supernova to the host galaxy's center.

We have automated this process with a gradient ascent method able to accurately locate the center of the host galaxy in which a cataclysmic event is embedded. We release this code with the rest of the \texttt{GHOST} software. 

For each transient, we have downloaded an 800x800 pixel (200$\arcsec$x200$\arcsec$) postage stamp of the field in $gri$. We have also downloaded the corresponding PS1 image stack masks and number counts. We use the stack masks to remove saturated pixels, and the counts mask to remove pixels created from fewer than two stack images. To remove stars that have not saturated the detector and any remaining starlight from those that have, we use the \texttt{DAOStarFinder} routine within the \texttt{photutils} package. We identify and remove sources matching a gaussian light profile with FWHM$\sim$3 at a threshold of 20 standard deviations above the median pixel intensity. We have intentionally set our threshold high to prevent us from masking AGN at the centers of galaxies, as these provide meaningful gradient information. We then remove each identified star using a circular mask with a 5 pixel radius. 

Once we have removed stars from the field, we use \texttt{Scipy} to construct a two-dimensional cubic spline interpolation over the pixels that were previously masked. We apply zero-points to these $gri$ images, average them together in magnitude space, and convert back to flux to generate a PS1-averaged image $A$. Next, we use the \texttt{Background2D} function within \texttt{photutils} to create an image dominated by the light profiles of galaxies within the field. This function is typically used to estimate spatially-varying background noise from a crowded field and study smaller sources in detail. Because a galaxy halo can be conceptualized as background noise to the structure within the halo, we can use this same function to create a version of image $A$ dominated by the light profiles of galaxies in the field and with minimal remaining effects from dust and HII regions. This smoothed image, which we call $B$, forms the basis for our gradient ascent method. 

To estimate the sky background, \texttt{Background2D} divides a pointing into evenly sized smaller regions, applies a median filter across each smaller region, then interpolates the resulting pixels back to the resolution of the original postage stamp. This smoothing requires a careful selection of both the size of each sub-region, or box, and the size of the median filter. Ideally, the size of each box should be larger than the galactic structure comprising the host galaxy of interest, and small enough to capture its radial light profile. A large sub-region and median filter will smooth local brightness variations; a small sub-region and median filter will preserve them. As a result, the size of the sub-region needed to resolve a host galaxy's light profile will change drastically from field to field. We have empirically determined a set of criteria for predicting whether the true host galaxy will fall into one of three size categories - ``small", ``medium", or ``large". These criteria determine the sizes of the image boxes and the median filter. A box of 75x75 pixels is used for large host galaxies, while boxes of 40x40 and 15x15 px are used for medium and small host galaxies, respectively. The median filter used for large and medium host galaxies is 3x3 pixels; for small host galaxies, no filter is applied to the data. 

The estimated size of the true host galaxy is based upon four measurements:

\begin{enumerate}
    \item $\tilde{I}$: The sigma-clipped median count of the image. This metric attempts to characterize the flux of non-stellar light across the full postage stamp. A high value of $\tilde{I}$ suggests a large host galaxy in the field, but can also indicate bright stars that were not completely removed. 
    \item $f_{IM}$: The fraction of pixels in the 800x800 pixel image with a photon count of at least unity. This metric provides additional evidence for a single extended source dominating the field, and is less biased by stellar saturation. 
    \item $f_{SN}$: The fraction of pixels within a 200x200 pixel box centered on the supernova with a photon count of at least unity. This provides strong evidence for the spatial extent of the true host galaxy. If both $f_{SN}$ and $f_{IM}$ are near unity, a host galaxy is deemed large; if $f_{SN}$ is large but $f_{IM}$ is small, a medium host galaxy is assumed.  
    \item $I_{SN}$ : The image count at the location of the supernova. If this value is above a certain threshold, the host galaxy is deemed ``small". This may appear counterintuitive, but this designation suggests that the supernova occurred near its host galaxy center. It is important that we use a small filter in this case to preserve the precise location of the host galaxy center.
\end{enumerate}

A large filter can shift the location of a host galaxy center due to a combination of AGN masking and edge effects. To recover the local peak, we normalize and combine images $A$ and $B$ into a weighted average (image $C$). The weights applied to $A$ and $B$ are determined by the magnitude of $I_{SN}$ and the number of masked pixels in image $A$. Image $B$ is given the dominant weighting in the resulting image unless $I_{SN}$ is above a threshold, for the same reason described above. If more than $15\%$ of the image has been masked, the cubic interpolation of the image becomes unreliable and artificial photon counts from interpolated masked stars contaminate $A$. If this occurs, the reference image $A$ is given a weight of $0$ and $C = B$.

Once we have our final profile-dominated image $C$, we estimate the two-dimensional gradient at each pixel using the \texttt{numpy} package. We use second-order central differencing at the interior points of each image and first-order differencing at the boundaries. We begin at the location of our supernova $(x_i, y_i) = (x_{SN}, y_{SN})$ in pixels and update the position according to the image gradient $\nabla C(x_i, y_i)$, using a forward Euler updating scheme: 
\begin{equation}
    (x_{i+i}, y_{i+1}) = (x_{i}, y_{i}) + h\nabla C(x_i, y_i)
\end{equation}
Here $h$ represents a chosen step size. After completing a step, we round our updated position coordinates to the nearest pixel. By tracing the local gradient in the postage stamp, the algorithm ``crawls" to local areas of increased brightness. For spatially extended sources, the gradients will cause the step to terminate at or near the center of the supernova's host galaxy. The algorithm iterates until the calculated position has reached the edge of the postage stamp or for 1000 steps, whichever comes sooner. If the algorithm reaches a region of the image with a gradient smaller than a pixel, we use the \texttt{numpy} \texttt{random} function to randomly perturb its location by one pixel in any direction. This feature, combined with the large number of total iterations, prevents the algorithm from getting trapped in a local maximum. After the algorithm terminates, we conduct a PS1 cone search at the terminated position. The radius of this search is 5$\arcsec$ for ``small" or ``medium" host galaxies (see above criteria), and 20$\arcsec$ for ``large" host galaxies. We eliminate sources with only one PS1 detection, and then select the non-stellar source closest to the terminated position. We illustrate this algorithm in Figures \ref{fig:background_SN2017erp} and \ref{fig:quiverDLR_SN2017erp}.

For this algorithm to be successful, we must select a physically relevant step size $h$. This must be large enough to ``overlook" remaining deviations from a smooth light profile, and small enough to prevent overshooting the true center of a host galaxy. For a given field, We calculate $h$ by multiplying the mean Kron radius for all potential host galaxies by an empirical scale factor based on the four criteria outlined above. We use a large scale factor when the supernova occurs within a large galaxy dominating the postage stamp, and a small scale factor when the supernova occurs within a small extended source. This allows our step sizes to vary with the scale of the most likely host galaxy even when the true host galaxy is unknown. This scale factor also accounts for unrealistically low PS1 Kron radius values, as is described in greater detail in \ref{Methods:PS1Issues}. If we adopt a scale factor of unity, the algorithm terminates in a majority of fields before the galactic center is reached.

%An advantage of this method is that it is able to find the center of a host even if the light profile is not axisymmetric about the face of the host galaxy. Supernovae located in the arms of spirals can also be associated with this method so long as the light profile of the arms increase toward the galactic center, as is illustrated in Figure \ref{fig:PSNJ16545668+4303077}. 

 %While the majority of galaxies do not have a light profile that increases monotonically toward the galactic center, we have found that we are able to reassociate most hosts using the estimated background derived from the methodology outlined above. 
\begin{table*}[t]
  \centering
%\begin{tabular}{ccc} \\ \hline
\begin{tabular*}{\textwidth}{ccc}
    Pipeline Step & Fraction Removed & Total Fraction \\ \hline \hline
PS1 30$\arcsec$ cone search for nearby host galaxies & 7.3\% & 92.7\% \\
Removal of sources not detected in $gri$ & 3.2\%  & 89.5\% \\
Removal of sources with \verb|qualityFlag = 128|, \verb|primaryDetection = 0|, \verb|bestDetection = 0| & $<$0.1\%&  89.5\%  \\ 
Star, galaxy separation with SVM & 3.6\% & 85.9\%  \\
Directional light radius host galaxy association & 10.3\% & 75.6\%  \\
Gradient ascent host galaxy association & -3.9\% & 79.5\%  \\  Redshift mismatch & 1.9\% & 77.6\%  \\
Manual reassociation & -0.4\% & 78.0\%  \\ \hline
    \end{tabular*}
    \caption{The fraction of total spectroscopically classified supernovae removed from our database at each stage of the association pipeline. We also list the total SN fraction remaining after each step.\label{tab:Cuts}}
\end{table*} 

 We trigger this algorithm when a host galaxy is not found using DLR, and when a candidate host galaxy is missing the morphology information necessary for the DLR method. There are 2,137 supernovae matching the first criterion, and 3,129 matching the second. We find by visual inspection that the gradient ascent method achieves a $12\%$ misassociation rate for the incomplete morphology sample, compared to $16\%$ for the DLR method. We additionally find a $\sim3\%$ misassociation rate for the ``hostless" sample, which generally consists of less crowded fields. More importantly, our method recovers $800$ host galaxies from the hostless sample. This is nearly half of the supernovae removed by the DLR method and roughly $4\%$ of all spectroscopic supernovae. We have included these matches in our final database. 

Our gradient ascent method reliably recovers transient host galaxies at low-redshifts when a large galaxy dominates the field, and at high-redshifts with sparse fields. When the host galaxy is large and the field is crowded, the large filter selected can blend multiple galaxies and shift the peak of the true host galaxy's light profile. This causes the algorithm to terminate at an artificial maximum, and can lead to incorrect associations. 

To be useful for early-time classification, our host association pipeline must be fast. We have timed our complete pipeline for five low-$z$ ($z < 0.001$) events and five high-$z$ ($z > 0.4$) events from TNS. We find a median of $85 \pm 4$ s per event for the low-$z$ sample and a median of $73 \pm 37$ s per event for the high-$z$ sample. The large spread in association times at high-$z$ is caused by the gradient ascent method, which is only triggered when the DLR method fails. More low-$z$ events triggered gradient ascent because the large physical projections involved makes it harder to find the center of the true host galaxy. The majority of time elapsed during the gradient ascent method is spent querying PS1 for postage stamps of the field; if images are provided in advance (or the association is done over a faster internet connection), latency will likely decrease.

\begin{figure*}[!ht]
    %\captionsetup[subfigure]{labelformat=empty}
     \centering
     \subfloat[][]{\includegraphics[width=0.5\linewidth]{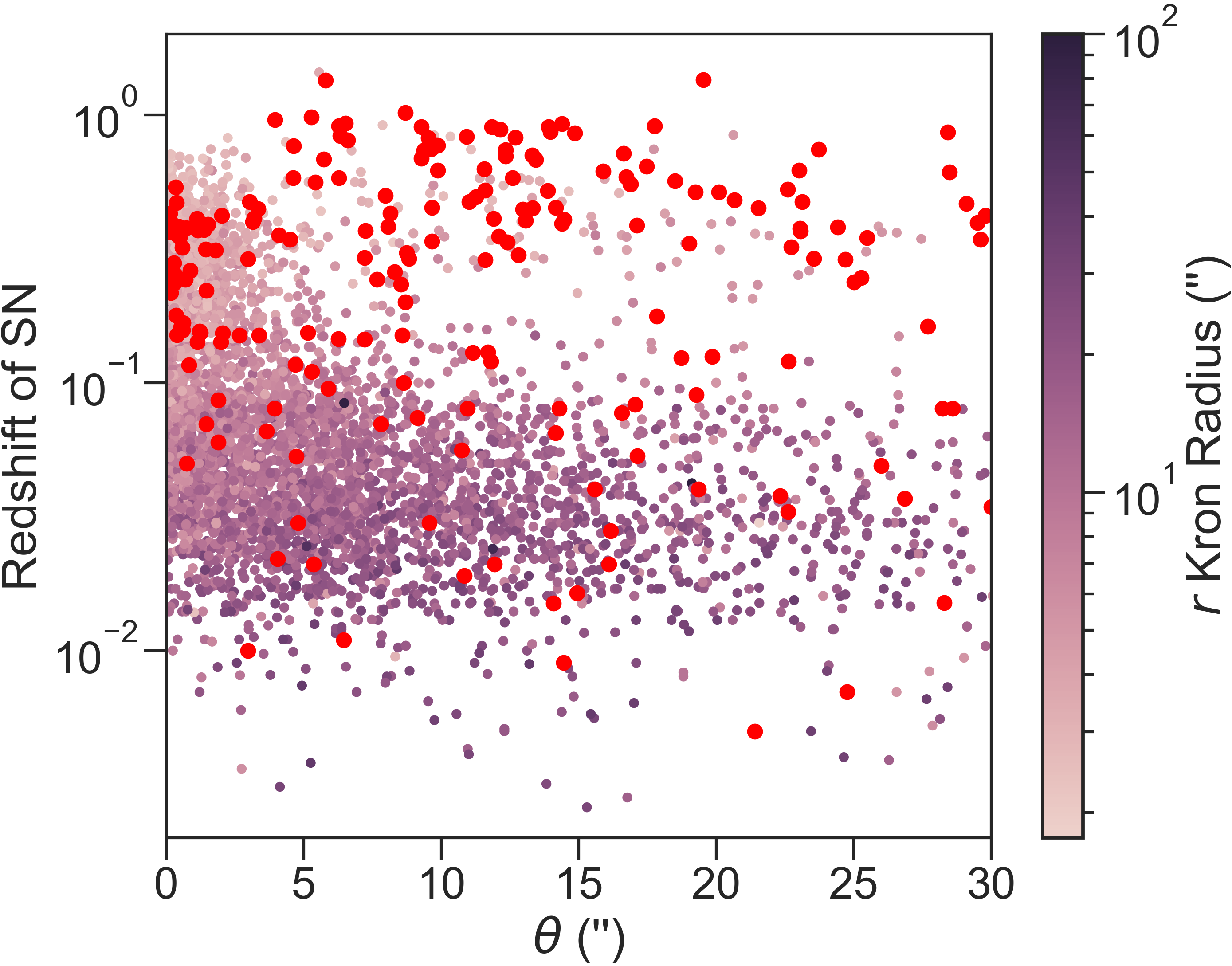}\label{fig:Redshift_mismatch}}
     \subfloat[][]{\includegraphics[width=0.5\linewidth]{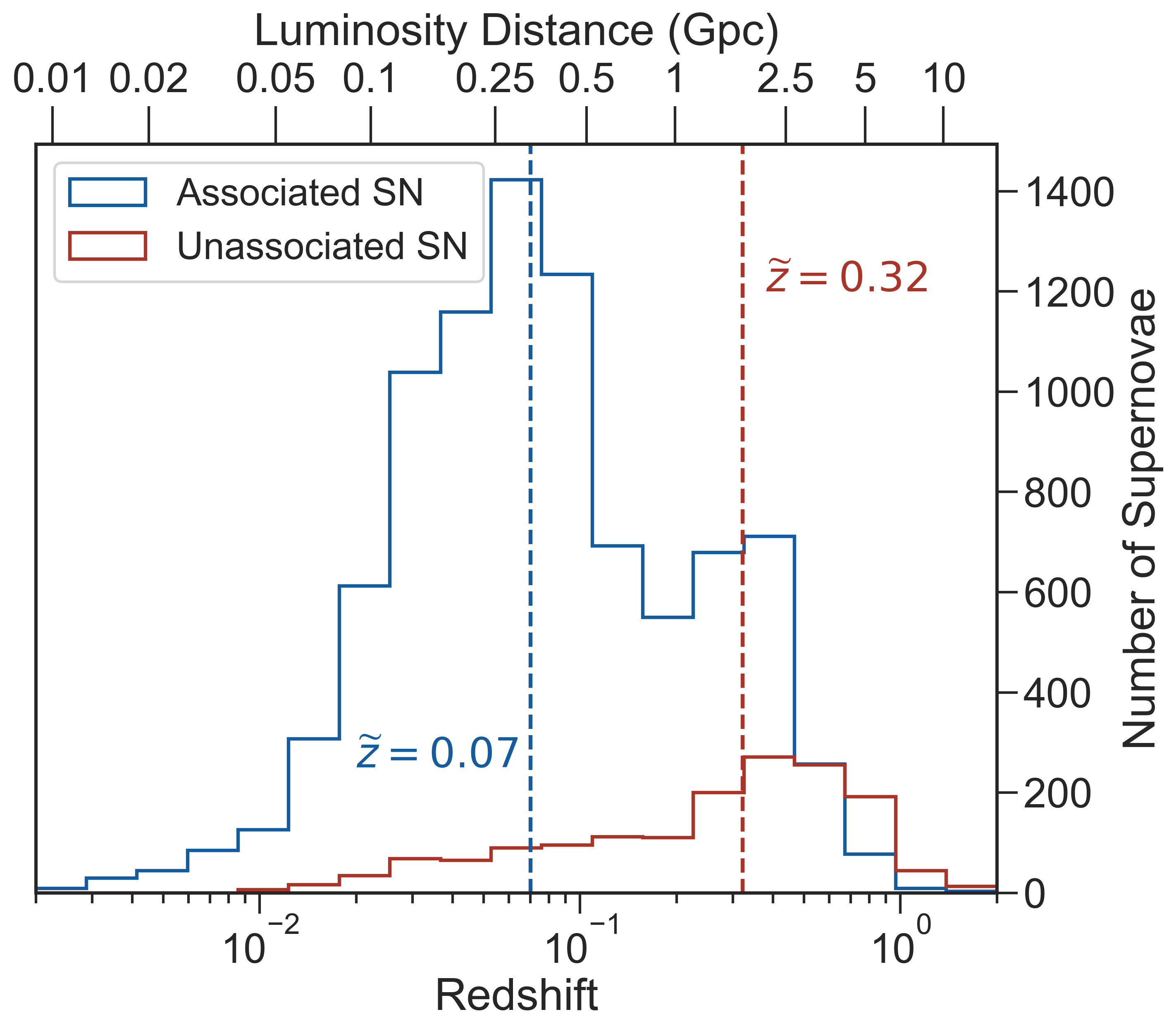}\label{fig:unassociated_redshifts}}
     %\vspace*{-5mm}
     \caption{\textbf{a.} The redshifts of supernovae in our sample as a function of angular separation from their host galaxy. Small points are colored by the host galaxy's $r$-band Kron radius. We find a $\theta$ distribution roughly matching that given in \cite{galbany2012type} but with multiple high-$z$ host galaxies with $\theta > 15\arcsec$. In most of these high-$z$ cases, the host galaxy's Kron radius is too small to reasonably explain this wide separation. By flagging the systems where the host galaxy's redshift differs by more than 50\% of the supernova's redshift (the large points in red), we confirm that this high-$z$ sample is composed of predominantly misassociated systems.} \textbf{b.} The distribution of spectroscopic redshifts for supernovae with associated host galaxies in \texttt{GHOST} and those without identified host galaxies. The median of each distribution is shown. Assuming a standard cosmology (H$_0$=67.8, $\Omega_0 = 0.3$) and a flat Universe, the supernovae in our sample lie at a median luminosity distance of $\approx$318 Mpc, while the unassociated sample lies at a median luminosity distance of $\approx$1.7 Gpc. We are able to associate the majority of supernovae within $z < 0.62$; the supernova sample analyzed in \cite{gupta2013understanding} spanned the range $0.01 \leq z \leq 0.46$.
\end{figure*}

 After incorporating host galaxies matched with the gradient ascent method, we manually re-associate an additional 425 supernovae ($\sim2.6\%$ of our table) by eye that were mismatched in our sample and another 69 low-$z$ supernovae ($0.4\%$ of our table). These low-$z$ supernovae were the events from TNS with $z<0.014$ ($d_L \approx 60$ Mpc) that were not successfully associated by our pipeline. In the majority of cases, the hosts of these supernovae were either $>$30$\arcsec$ from the event or reported unphysical radius information in PS1 and dropped at the DLR stage. We further identify misassociations by plotting the redshift of each supernova as a function of its angular separation $\theta$ from its host galaxy. Multiple associations indicate $\theta > 15\arcsec$ and $z > 0.1$, corresponding to a physical separation of $>30$ kpc. Although supernovae have been discovered as far as 80 kpc from their host centers from tidal interactions with other galaxies \citep{ferretti2017probing}, events at this separation are probably uncommon (see Figure 2 of \citealt{galbany2012type}). To confirm this sample as misassociated, we identify the matches where the supernova redshift differs from its host galaxy's redshift by greater than 50\%. We plot this sample in Figure \ref{fig:Redshift_mismatch}. These $\sim300$ events occur predominantly at $z_{SN} > 0.1$ and comprise the majority of our high-$z$ matches with large separation. We have removed these events from our database. We estimate the misassociation rate of our final database to be $<3\%$, lower than the missasociation rate reported by \cite{gupta2016host}. We list the fraction of total supernovae dropped from our sample at each stage of the pipeline in Table \ref{tab:Cuts}. 
 
 Our final database contains host galaxies for $78\%$ of all spectroscopically classified supernovae, higher than the completeness reported by \cite{gupta2013understanding} (73\%, or 15,826/21,787 candidates) and spanning a significantly wider range in redshift. We have associated the majority of spectroscopic supernovae within $z<0.62$; the furthest supernova candidate in \cite{gupta2013understanding} was at $z = 0.46$. 33.2\% of events within our unassociated sample occurred at declinations below $-30\degree$, where a host galaxy search was excluded because no PanSTARRS 3PI survey images would be available. Excluding these events brings the fraction of associated supernovae to 84\%. The remaining unassociated supernovae are located at significantly higher redshifts than the supernovae in the associated sample, as is shown in Fig. \ref{fig:unassociated_redshifts}. It is possible that even PS1 depths are insufficient to resolve these host galaxies.
 
$43$ supernova pairs in our final table have the same redshift, similar discovery dates (within 100 days of each other), and coordinates within $1$ arcsecond of each other. These duplicates were removed. Two supernovae within $1$ arcsecond of each other have discovery dates greater than one year apart. This pair of events is SN007ie, a type Ia supernova discovered on May 9th, 2007 at an RA and Dec (J2000) of 334.4029$^{\circ}$, 0.6133$^{\circ}$ ; and PS1-11aqj, a type II supernova discovered on February 9th, 2011 at an RA and Dec of 334.4028$^{\circ}$, 0.6134$^{\circ}$.
    %\newpage  
    \section{Data and Software Products}\label{softwareSection}
    \subsection{The GHOST Database\footnote{\href{https://github.com/uiucsn/astro_ghost}{https://github.com/uiucsn/astro\_ghost}}} 
    The Galaxies HOsting Supernova Transients (\texttt{GHOST}) database contains \GHOSTcount spectroscopically classified supernovae and the photometric properties of their host galaxies. The full table can be downloaded as a CSV, and the host galaxy data for OSC supernovae have been added to the JSON data files and re-released for convenience. We have also included 1,436 masses estimated from host galaxy photometry (Foundation, \citealt{foley2018foundation}; PS1COSMO, \citealt{scolnic2018pantheon}, \citealt{2018JonesSample}), Hubble residuals of 124 SNe Ia (\texttt{Kaepora}, \citealt{siebert2019investigating}), host galaxy and transient spectra (Sloan Digital Sky Survey, \citealt{york2000sloan}; TNS), and transient light curves. In addition, we have developed software for querying the database and associating new host galaxies. We describe the functionality of the code in the following sections.
%    Supernova spectra have been scraped from TNS and host galaxy spectra are taken from the Sloan Digital Sky Survey (SDSS; \citealt{york2000sloan}) and NED. 

       \begin{figure*}
        \centering
        \includegraphics[width=\linewidth]{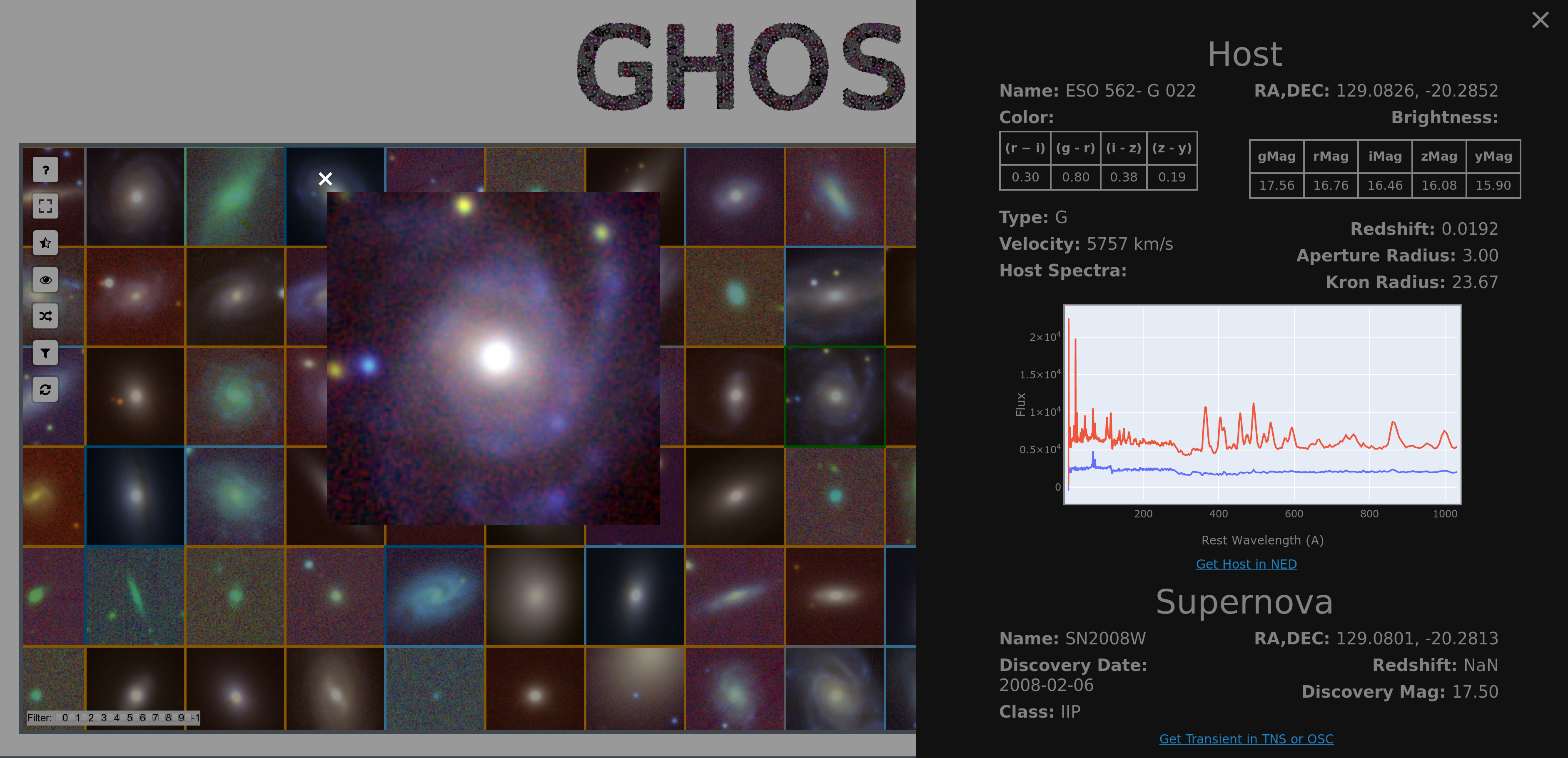}
        \caption{A screenshot of the \texttt{GHOST} Viewer, which consolidates \GHOSTcount host galaxy postage stamps and basic information about the supernovae they host into a streamlined GUI. A pop-up sidebar provides spectra for both supernova and host galaxy, if they are available. Light curve data from the Open Supernova Catalog is also presented. By viewing many identified host galaxies simultaneously, incorrect associations can be rapidly identified and population statistics can be studied in greater detail.}
        \label{fig:website}
        \end{figure*}
        
    \subsection{Analysis Tools for the GHOST Database}\label{software}
        We have developed and released a Python package called \texttt{astro\_ghost} to analyze the \texttt{GHOST} data products. It can be found at the repository for this work. The package also contains our full DLR and gradient ascent association pipelines. We list several functions for parsing the database below, along with a basic description of their use:
    
        \begin{itemize}
          %  \item \verb|getHostFromTransientCoords|: Inputs the location of a transient as an \texttt{astropy SkyCoord} object, and returns the host galaxy associated with that transient (if it exists in the database).
          %  \item \verb|getHostFromTransientName|: Inputs the TNS name of a transient, and returns the associated host galaxy (if it exists in the database).
           \item \textsf{getTransientHosts}: Inputs the name and location of a transient or list of transients. If the transient is not found in the database by name or location, the association pipeline is run and the most likely host galaxy (one per transient) is returned.
            \item
            \textsf{getHostStatsFromTransientCoords}: Inputs the location of the transient as an \texttt{astropy SkyCoord} object, and returns basic statistics about a host galaxy (including other associated transients in the database).
            \item
            \textsf{getHostStatsFromTransientName}: Inputs the TNS name of a transient, and returns basic statistics about a host galaxy.
            \item \textsf{getTransientStatsFromHostName}: Generates basic statistics for a transient (or a series of transients) based on the NED-reported name of its host galaxy.
            \item \textsf{getTransientStatsFromHostCoords}: Generates basic statistics for a transient, based on host galaxy location as a \texttt{astropy SkyCoord} object.
            \item \textsf{getHostImage}: Inputs a transient TNS name and returns a postage stamp of the most likely host galaxy in one of the PS1 bands - $g,r,i,z,y$ - as a fits file, and plots the image in $gri$ as a color image. 
            \item \textsf{getTransientSpectra}: Plots the spectrum of the transient from TNS, if it is available. 
            \item \textsf{getHostSpectra}: Plots the spectrum of the host galaxy from SDSS or NED.
            \item \textsf{coneSearchPairs}: Completes a cone search for all transient-host pairs whose transient location falls within a certain radius, returned as a pandas dataframe. Useful for identifying supernovae associated with the same system where the NED name in the database is that of an HII region or star within the galaxy.
            \item \textsf{fullData}: Returns the full \texttt{GHOST} database.
        \end{itemize}

        Sample code outlining the usage of each of these functions is provided at the link above.
        
        \subsection{The GHOST Viewer\footnote{\href{GHOST.ncsa.illinois.edu}{GHOST.ncsa.illinois.edu}}} 
        We have also created a website for simultaneously viewing all host postage stamps in our database. This website is hosted at the National Center for Supercomputing Applications (NCSA). Postage stamps are dynamically scaled in real time so that the user may rapidly view many host galaxies or individual systems of interest. Host galaxy images are color-coded by the class of the supernova matched to them. After selecting a postage stamp, a basic summary of both the supernova and its host galaxy are provided in a sidebar. Users may interactively search the database for a specific supernova by name, spectral class, or by the name of its host galaxy. Interactive plots showing the spectra of both host galaxy and supernova are also presented, and this data can be downloaded as a CSV. Both \texttt{GHOST} and its associated viewer will be updated as new supernovae are reported. 
        % The GHOST Viewer allows the user to query the database by supernova or galaxy name, and the postage stamps can be further filtered by supernova class.

\section{Dimensionality Reduction}\label{DimensionReduction}
Our final dataset contains 317 features of \GHOSTcount PS1 sources, along with the following seven features of their associated supernovae: right ascension, declination, class, name, redshift, angular offset $\theta$, and scaled angular offset $\theta/d_{DLR}$. Nearly half of the host galaxy features characterize the source's detection in PS1. These include the pixel coordinates of the source in each filter, the number of source detections, and the detection ID in each band. Nevertheless, the large number of remaining features prohibits a brute-force search for supernova correlations. Further, galaxy features are highly correlated with each other, leading to empirical relationships such as the color-magnitude relation \citep{bell2004nearly} and the Fundamental Plane for ellipticals \citep{dressler1987spectroscopy}. These properties can reveal similar information about a galaxy's position along an evolutionary track. As a result, these features likely do not provide unique information about the class of a supernova.

To test this hypothesis, we construct a correlation matrix, given in Figure \ref{fig:g_Corr}, for the galaxy features in \texttt{GHOST}. We use Spearman's rank correlation over the Pearson correlation because the latter describes only linear relationships, whereas the former quantifies the ability of a relationship to be described by any monotonic function. For features present in each band, we have only shown those features in $g$. We find that over half of our single-band features are $>80\%$ correlated, and the same features across multiple bands are even more strongly correlated. The strong correlations between magnitude, flux, and radial moments of host galaxies in \texttt{GHOST} appear as blocks within the full correlation matrix. These blocks form a regular grid across our matrix due to the strong correlations between brightness features in each band. Aperture radius is not strongly correlated with our other photometric features, and therefore appears as a gap in the correlation blocks.

 \subsection{Identifying Dominant Host Galaxy Features with Principal Component Analysis}\label{PCA}
 
   \begin{figure*}
    \centering
    \includegraphics[width=\linewidth]{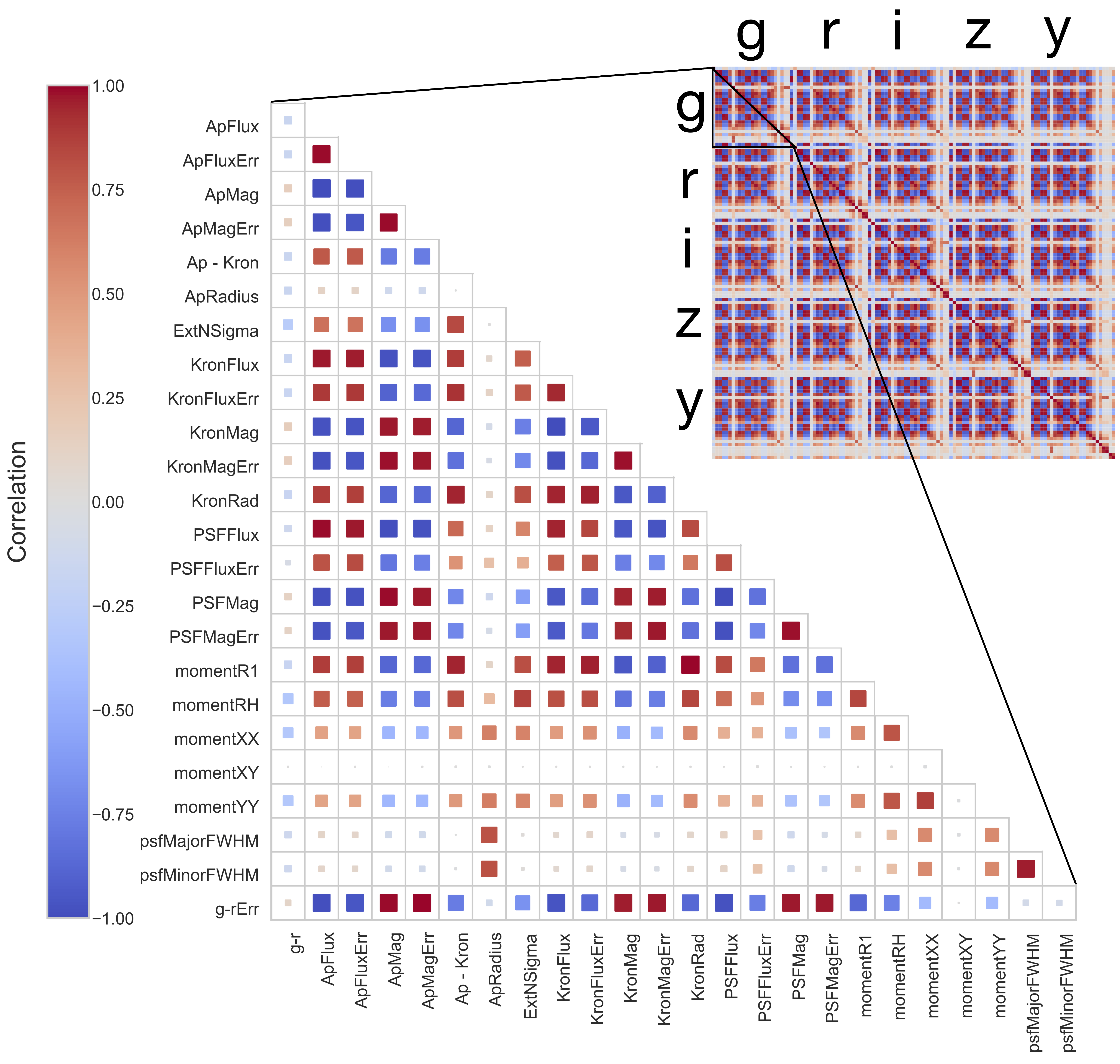}
    \caption{The Spearman rank correlation between PS1 source features in a single band (\textit{g}), with blue corresponding to negatively correlated features and red corresponding to positively correlated features. Over half of the features are strongly correlated, particularly the magnitude of the host galaxy in different apertures and its first and second order radial moments. Strong correlations can also be seen in the same features across multiple bands (inset at right), leading to the block structure of the matrix.}
    % Because this matrix describes both linear and non-linear correlations, it is able to capture the relationship between flux and magnitude.
    \label{fig:g_Corr}
\end{figure*}

To reduce our list of host galaxy parameters without reducing their predictive power, we undertake a principal component analysis (PCA) of our galaxy data. In PCA, a dataset is transformed to a set of orthogonal components with the first principal component containing the most uncorrelated variance. A dataset composed of highly correlated features will be well-characterized by a single principal component, which is a linear combination of the original features. 

We begin by re-scaling our host galaxy features so that each has a mean of $0$ and a variance of $1$. Each of our features is continuous instead of categorical, rendering one-hot encoding unnecessary. Re-scaling is necessary to directly compare features whose values may differ in both magnitude and range. If a single feature contains a wide range of values, the first calculated principal component will capture most of the variance in this feature alone and not across all features. Next, we transform our \texttt{GHOST} database to its first two principal components using the \texttt{sklearn} package in Python. These principal components capture over half ($55\%$) of the total variation in host galaxy features, confirming the degeneracy shown in Fig. \ref{fig:g_Corr}. 

\begin{figure}
    \centering
    \includegraphics[width=\linewidth]{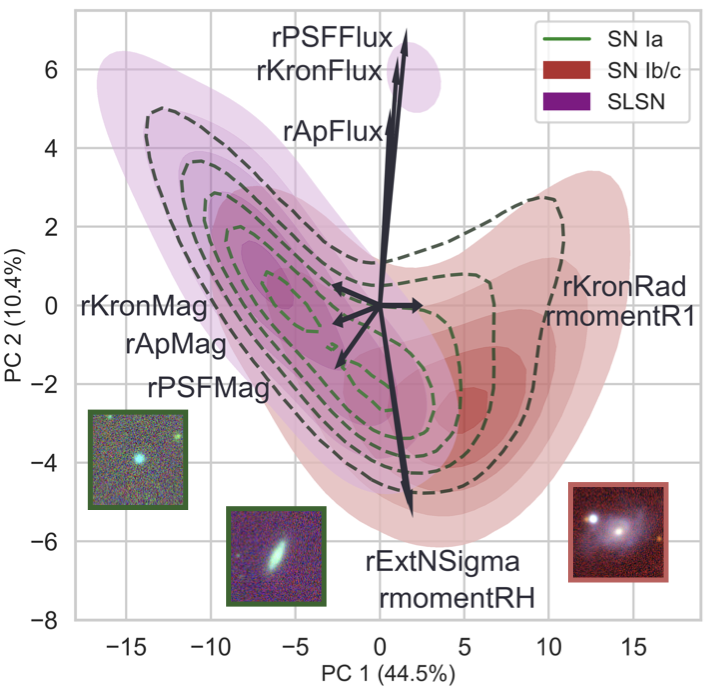}
    \caption{The PCA-transformed distribution of host galaxy parameters for three supernova classes in our data: SN Ia, SN Ib/c, and SLSN.%, transformed into our reduced principal component space. 
    For clarity, we plot in black the feature vectors in $r$ with highest loading along the two principal component axes. The SN Ib/c and SLSN distributions are centered at different positions along the first principal component, suggesting host galaxies with different size and brightness properties. The distribution of SN Ia host galaxies is bimodal, suggesting multiple host galaxy populations. A distinct group of SLSNe can also be seen top center. These galaxies host unusually energetic SNe (see text) -- an indication that SN anomalies can be found by the properties of their host galaxies. Postage stamps correspond to the peak of the SN Ib/c distribution (orange border) and the peaks of the SN Ia distribution (green borders)}.
    \label{fig:PCA_2Comp}
\end{figure}

By reducing the dimensionality of our dataset, we can visually identify differences between classes of supernovae. We present a biplot of our principal component model in Figure \ref{fig:PCA_2Comp} along with the transformed distributions of SN Ia, SN Ib/c, and SLSN host galaxies. Because the principal components are linear combinations of the galaxy features, we can represent the initial galaxy features as vectors in this reduced space. The loading of each vector is defined as its projected length along each PCA axis; this quantifies the weight assigned to that feature in calculating a principal component. Highly linearly correlated features are given similar weights along the two axes, and we can see a clustering of magnitude, flux, and morphological features forming the dominant contributions to this PCA space. The first principal component, which explains $44.5\%$ of the variation in host galaxy information, is determined nearly equally by the Kron radius of a source and its PSF, Kron, and Aperture magnitudes. The second principal component, which explains $10.4\%$ of the variation, is determined mainly by the PSF flux, $momentRH$ (the square root of the first radial moment) and $ExtNSigma$ in each band (a PS1 derived feature extremely similar to our $m_{\rm Ap} - m_{\rm Kron}$ mag, $ExtNSigma$ is defined as the difference between PSF and Kron magnitude of a source subtracted by the stellar median value and divided by the error combined in quadrature; see \citealt{magnier2016pan} for more information). SN Ia host galaxies are widely distributed in this space, with two peaks at separate locations along both principal components. The two highest-density contours for host galaxies of SNe Ib/c span a significantly smaller range in PCA space than the highest-density contours for host galaxies of SNe Ia, although with a smaller sample size (591 SNe Ib/c vs. 6,279 SNe Ia) this homogeneity may be unphysical. These three classes are also centered at different locations in PC space, providing a useful diagnostic tool for distinguishing classes of supernovae without any information from their explosions.

The multimodal distribution of SN Ia host galaxies in this PCA space suggests multiple distinct populations, particularly given the large sample size of this class. In addition, the strong overlap with the two other groups suggests that SN Ia host galaxies are photometrically similar to galaxies hosting SNe Ib/c and SLSNe. By embedding postage stamps corresponding to each peak in Fig. \ref{fig:PCA_2Comp}, we can see that one SN Ia mode aligns with bright spiral galaxies (which are similar to SN Ib/c hosts), whereas the other aligns with faint, featureless hosts (which are similar to SLSN hosts). The bimodality of SN Ia host galaxies also presents a challenge for classification. SN Ia residing within host galaxies at the left peak are likely to be misclassified as SLSNe, and SNe Ia within host galaxies at the right peak are likely to be misclassified as SNe Ib/c. Despite this overlap, the separation along the first principal component suggests that type information is revealed by host information alone.

By reducing the dimensionality of our host galaxy features, we can also rapidly identify outliers within \texttt{GHOST}. Fig. \ref{fig:PCA_2Comp} reveals a sub-sample of SLSN host galaxies that are well-separated from the rest of the distribution in PCA space. This population is comprised of four supernovae in our dataset. Two of these are SN2213-1745 and SN2016aps, events whose hyper-energetic explosions make them strong candidates for pair-instability or pulsational pair-instability supernovae (\citealt{cooke2012superluminous, nicholl2020extremely}). SN2016aps is the brightest supernova ever discovered, with peak absolute magnitude of $-22.35 \pm 0.09$ in $i$. Another supernova in this sample is SN2017gci, which was listed in the original AT report as a candidate cataclysmic variable; and LSQ14fxj, which has been alternatively classified as a SLSN, SN Ic, and an unusually bright Ia by different groups. From the feature loadings in Fig. \ref{fig:PCA_2Comp}, it is possible that some of these events correspond to the faintest SLSN host galaxies in our sample; indeed, SN2017gci was initially listed as hostless, and the estimated redshift of SN2213-1745 is $z=2.0458 \pm 0.0005$ \citep{cooke2012superluminous}, making it the most distant SN in our sample. The fact that these rare supernovae are associated with outliers in our host galaxy sample suggests a strong correlation between transient and host galaxy. Studying these host galaxies in more detail will likely shed light on these enigmatic events.

\begin{figure}
    \centering
    \includegraphics[width=\linewidth]{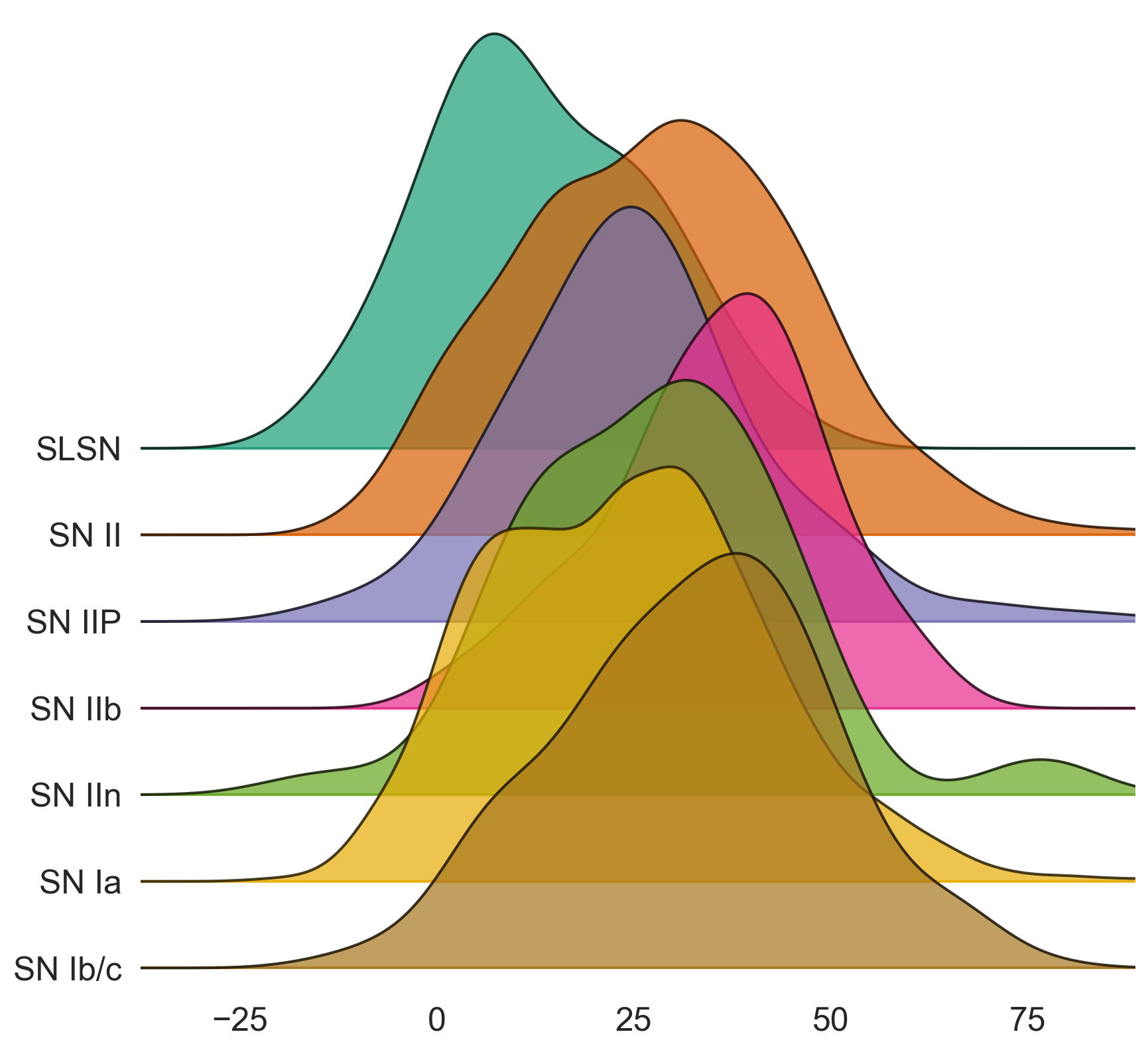}
    \caption{The distributions of supernova classes as a function of their loading along the first principal component from our PCA analysis. In this space, a type IIb supernova (SN IIb) is easily differentiated from a superluminous supernova (SLSN). The host properties of IIb, II, and Ib/c are nearly indistinguishable.}
    \label{fig:joyplot}
\end{figure}

Because significantly more variance from our full table is captured by the first PCA component, we can better distinguish SN classes by projecting the data along only this axis. We show Gaussian kernel density estimates (KDEs) for our SNe along the first principal component in Fig. \ref{fig:joyplot}. Because sample sizes vary dramatically between classes, the spread may not be representative of the true underlying populations; however, the medians of each distribution are illustrative. 

Broadly categorizing this principal component as ``size and brightness" and the second principal component as ``light profile"  in alignment with the loadings from Figure \ref{fig:PCA_2Comp}, we can now identify systematic physical differences between classes. The host galaxy distributions for type II, Ib/c, and IIb supernovae peak at nearly the same location and feature a heavy rightward skew, a strong indication that these supernovae are found in host galaxies with similar brightness and Kron radius. These hosts are distinct from SLSN host galaxies, which have a strong leftward skew; and host galaxies of Ia supernovae, which are not strongly skewed in either direction. By visually inspecting a subset of postage stamps corresponding to these populations, we find our SN Ib/c and SN II host galaxies to consist mainly of large spirals, whereas our SLSNe are hosted in smaller galaxies with a range of different morphologies. Visual inspection of our SN Ia host galaxy sample did not reveal any consistent trends with respect to size or morphology, potentially explaining the wide spread of the SN Ia KDE. This suggests that the horizontal axis in Figure \ref{fig:joyplot} corresponds roughly to brighter, larger galaxies toward the right and smaller, fainter galaxies toward the left. These results are in agreement with previous findings that SLSNe-I are found almost exclusively in low-mass dwarf galaxies with low star formation rates (\citealt{perley2016host, 2015Leloudas, 2015Lunnan, 2016Angus}) and core-collapse supernovae occur predominantly in late-type galaxies with high star formation rates (\citealt{Cappellaro1999, Hamuy2003}). 

The multiple peaks of SNe Ia host galaxies from Fig. \ref{fig:PCA_2Comp} are also seen in Fig. \ref{fig:joyplot}. This provides additional evidence for multiple host galaxy populations with distinct size, brightness, and morphology. Previous studies have shown that SNe Ia occur within a broad range of galaxies \citep{de2006fundamental}. 

It is possible that these host galaxy differences can be attributed at least in part to differences in redshift. The fact that the postage stamp at the peak of our SN Ib/c distribution reveals significantly more morphological information than the postage stamp near the peak of the SLSN distribution seems to verify this prediction, as SLSNe are found at systematically higher redshifts than SNe Ib/c. In addition, the archival data we have used contain observational biases that systematically under-represent the number of supernovae at higher redshifts. While we are unable to completely constrain the redshift dependence of our features, no single magnitude-limited survey to date contains enough spectroscopically confirmed events of each class to construct a redshift-balanced sample. We have found our PCA results to be consistent for a sub-sample of our data consisting of only low-$z$ ($<0.014$) host galaxies, but we caution that this work represents only a first pass toward characterizing supernova host galaxies.
%\newpage

%\newpage
 \subsection{Comparing Host Galaxy Distributions with tSNE}\label{tSNEResults}
While PCA is valuable for visualizing data, it is only able to compose reduced dimensional spaces from linear combinations of features. This renders it unable to identify nonlinear correlations that may be valuable for our analysis. The limitations of this method can be seen in the wide separation between host galaxy magnitude and flux vectors in Fig. \ref{fig:PCA_2Comp}, suggesting no correlation. To explore nonlinear relationships between our host galaxy features, we implement t-Distributed Stochastic Neighbor Embedding (tSNE) \citep{maaten2008visualizing}. In tSNE, the transformation preserves the distribution of separations between points. This makes it another useful tool for directly comparing multiple classes of objects while preserving nonlinear correlations between features.

\begin{figure}
    \centering
    \includegraphics[width=\linewidth]{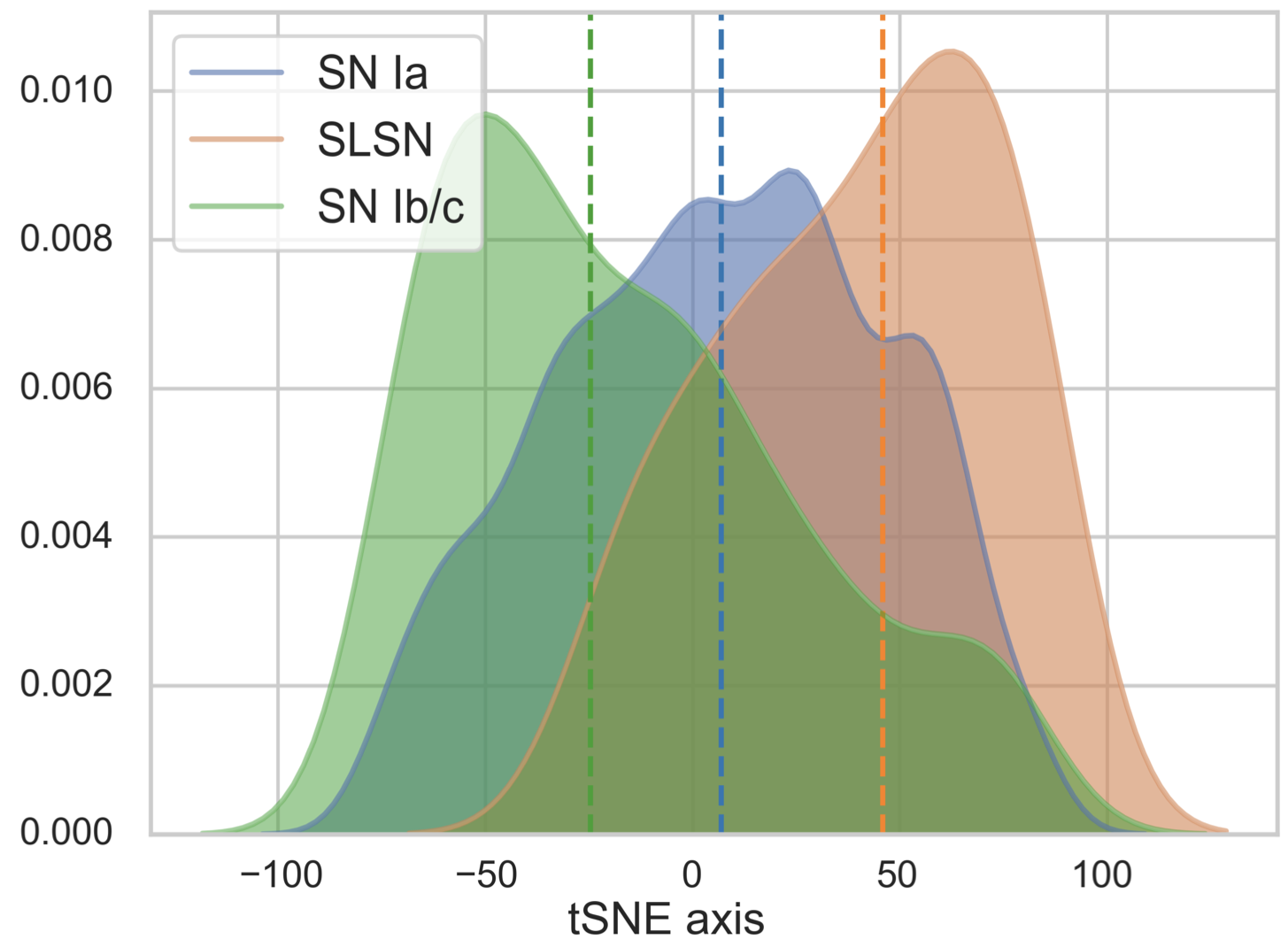}
    \caption{Distributions of host galaxy features for SNe Ia, SNe Ib/c, and SLSNe along the first dimension of our 2D tSNE, with dashed lines indicating the median of each distribution. Relative positions and shapes of the three distributions are consistent with the results from our principal component analysis given in Fig. \ref{fig:joyplot}, suggesting different underlying host galaxy populations.}
    \label{fig:tsne}
\end{figure}

We begin by projecting our data into a three-dimensional tSNE space using the \verb|sklearn| implementation. We select a perplexity of 30 and a learning rate of 200. To compare our results with the PCA analysis completed in \textsection \ref{PCA}, we project our supernovae along only the first tSNE axis. We present KDEs for our transformed host galaxy data for SNe Ia, SNe Ib/c, and SLSNe along this axis in Fig. \ref{fig:tsne}. As before, we see a leftward skew for SLSN host galaxies and rightward skew for SN Ib/c host galaxies. The medians of these distributions occur at significantly different positions along this axis. tSNE is a stochastic algorithm, and we have verified through multiple iterations that the differences between our distributions are insensitive to the random seed used. These results suggest that the systematic differences between core-collapse, SN Ia, and SLSN host galaxies are robust, with SLSNe preferentially found in small and faint host galaxies and core-collapse supernovae found preferentially in large, bright host galaxies. %where star formation is high. 

\begin{figure*}[!ht]
     \centering
      \subfloat[][]{\includegraphics[width=\linewidth]{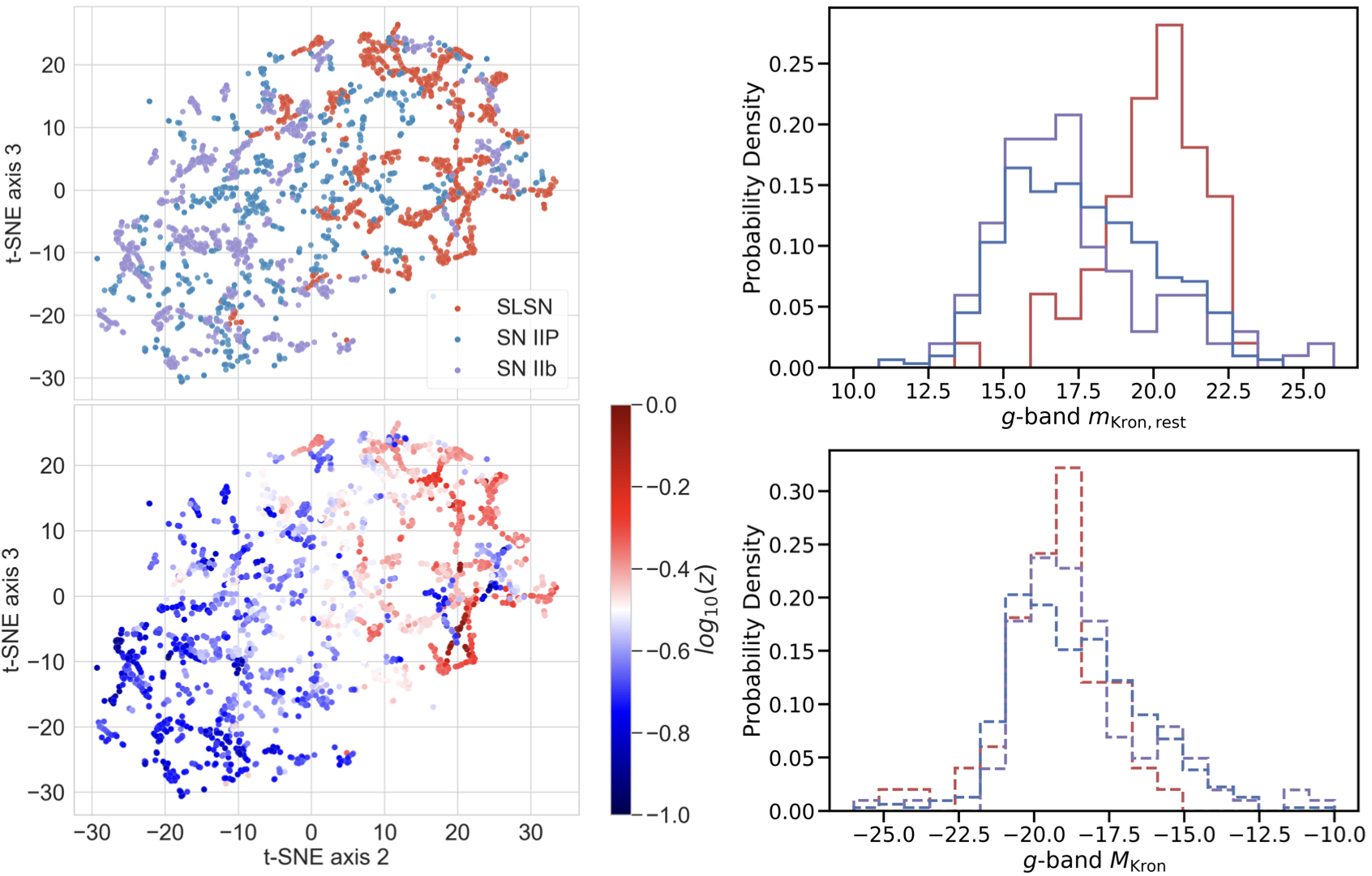}} 
    \caption{\textbf{Left, Above.} The distribution of host features corresponding to three classes of supernovae (SNe IIP, SNe IIb, and SLSNe) along our second and third tSNE axes. SLSN and SN IIb host galaxies are easily separable in this space, whereas SN IIP hosts are centered between the two distributions and span a much wider range of values. \textbf{Left, Below.} The same host galaxies as above, colored by redshift. The clustering of SNe IIb and SLSNe in tSNE space is highly redshift-dependent. \textbf{Right, Above.} Histograms for the k-corrected $g$-band Kron magnitudes of the same hosts as in left. Distribution colors match the supernova classes top left. SLSN host galaxies are fainter than nearly all SN IIb host galaxies; SN IIP host galaxies span both distributions. \textbf{Right, Below.} Histograms for the $g$-band absolute Kron magnitudes of these hosts. After correcting for host redshift, the distributions of the three classes are nearly coincident. This implies that the majority of the observed differences between SLSN, SN IIP, and SN IIb host galaxies are caused by redshift-dependent event rates.}
    \label{fig:2ComptSNE}
\end{figure*}

To further explore nonlinear correlations, we transform our host galaxy features into a three-dimensional t-SNE space. We find striking visual differences in the distributions of SN IIP, SLSN, and SN IIb host galaxies along the second and third t-SNE axes, which we plot in Fig. \ref{fig:2ComptSNE}. SLSN and SN IIb host galaxies are most easily distinguished in this space, and SN IIP hosts are centered near the middle of the two distributions and span the full space of SN host galaxy features. We have found the separation between SN IIb and SLSN host galaxies to be robust for a range of perplexities and initial random seeds, suggesting that real discriminatory information exists between these classes. These results also suggest that certain supernova sub-types may easier to distinguish using host galaxy features than SNe Ia and core-collapse events. Larger sample sizes from upcoming surveys, especially at higher redshifts, will make it possible to investigate the host galaxy correlations of these underrepresented events in greater detail.

Figure \ref{fig:PCA_2Comp} suggests that our ability to distinguish supernova host galaxies by class is redshift-dependent. This relationship becomes more evident when we plot the redshifts for the same SN IIP, SN IIb, and SLSN host galaxies described above. In the left panel of Figure \ref{fig:2ComptSNE}, we find that redshift roughly decreases toward the bottom left of our tSNE space and increases toward the top right. SLSNe events are highly coincident with high-redshift host galaxies and SN IIb events are highly coincident with low-redshift host galaxies. This reflects previous findings that SLSNe preferentially occur at high-$z$ and core-collapse events have been preferentially observed at low-$z$, particularly SNe IIb \citep{kelly2012core}. 

We can partially account for these redshift effects by calculating the absolute magnitudes of our host galaxies. First, we convert our PS1 magnitudes to the SDSS photometric system using the model developed by \cite{2016Finkbeiner}. We then k-correct our host galaxy magnitudes using the $SDSS$ color-conversion from \cite{2012Chilingarian} and convert to rest-frame using the galaxy redshifts found in NED. We show $g$-band rest-frame and absolute magnitudes of these hosts in the right panel of Fig. \ref{fig:2ComptSNE}. SN IIb and SLSN host galaxies in $g$ are well-separated by rest-frame magnitude, but after correcting for their redshifts reported in NED we find that the absolute magnitudes of these host galaxies are nearly identical. This supports the evidence from our 2D tSNE plot that previously discovered SNe IIb and SLSNe occur at distinct redshifts. Though these host galaxies are similar in intrinsic brightness, they can still be distinguished at early times by the redshifts at which they more frequently occur. A volume-limited survey would confirm whether these redshift-dependent rates reflect our observational biases or underlying formation mechanisms for SN IIb and SLSN events. If these trends persist in future datasets, they could also be used to validate photometric redshifts of these classes.

%validate rates of each class within simulated datasets. 
%This same validation can be applied in reverse, where a classification derived from spectral follow-up can inform photometric redshifts estimates for the supernova's host galaxy.

\section{Supernova Classification}\label{ClassificationMainSection}

\subsection{Random Forest: Methods}\label{randomforest}
%Having constructed a dataset of supernovae and the PS1 features of their host galaxies, 
We are now able to quantify the predictive power of host galaxy information in supernova classification. As in our star-galaxy separator, we implement a random forest model for this classification.

First, we remove all host galaxy features from our final table that do not describe physically meaningful information. These include PS1 IDs, PS1 data quality flags and features that reflect the quality of model fits. After removing rows containing missing values, we are left with 11,873 supernovae. In order to ensure accurate matches for classification, we also drop all rows where the redshift of the associated host differs by greater than 5\% from the redshift of the supernova (if both are reported). We then consolidate supernova class labels.

%In the second model, we consider exclusively core-collapse sub-types: SN Ib/c, SN IIP, SN IIb, and SN IIn. These sub-types are roughly balanced in sample size, and we have considered them following the null results of the IIP vs. IIn Anderson-Darling test in \textsection \ref{PCA} to investigate further whether these sub-classes truly originate in distinct host galaxy populations.}

First, we consolidate our training data into two classes: core-collapse supernovae (including SNe Ib/c, SNe II, and all sub-classes of SN II events) and type Ia supernovae (SNe Ia). The vast majority of supernovae within \texttt{GHOST} are SNe Ia, and in a magnitude-limited survey the majority of discovered events will also be SNe Ia. Training our random forest on the observed distribution of events would improve the classification accuracy when tested on data matching this distribution, but the tendency of the classifier to preferentially identify SNe Ia would prevent it from learning fundamental differences between the host galaxies of underrepresented supernova classes. The imbalanced training set would also lead to more wrong classifications for rarer events, as the algorithm would determine that any event with ambiguous host galaxy properties is likely a SN Ia. Further, the intrinsic rates of undersampled supernovae remain poorly understood \citep{strolger2015rate} and therefore poorly constrained \citep{prajs2017volumetric}. Observed rates also suffer from known systematic effects such as Malmquist bias \citep{leaman2011nearby}. Because we are unable to rebalance our dataset to match the intrinsic rates of each class, a model trained on this data would maximize accuracy by predicting on observed population rates. Because we are primarily focused on characterizing host galaxies and not event rates, we re-balance our training set using the package \texttt{Imbalanced-learn}.

\begin{figure}
     \centering
     \includegraphics[width=\linewidth]{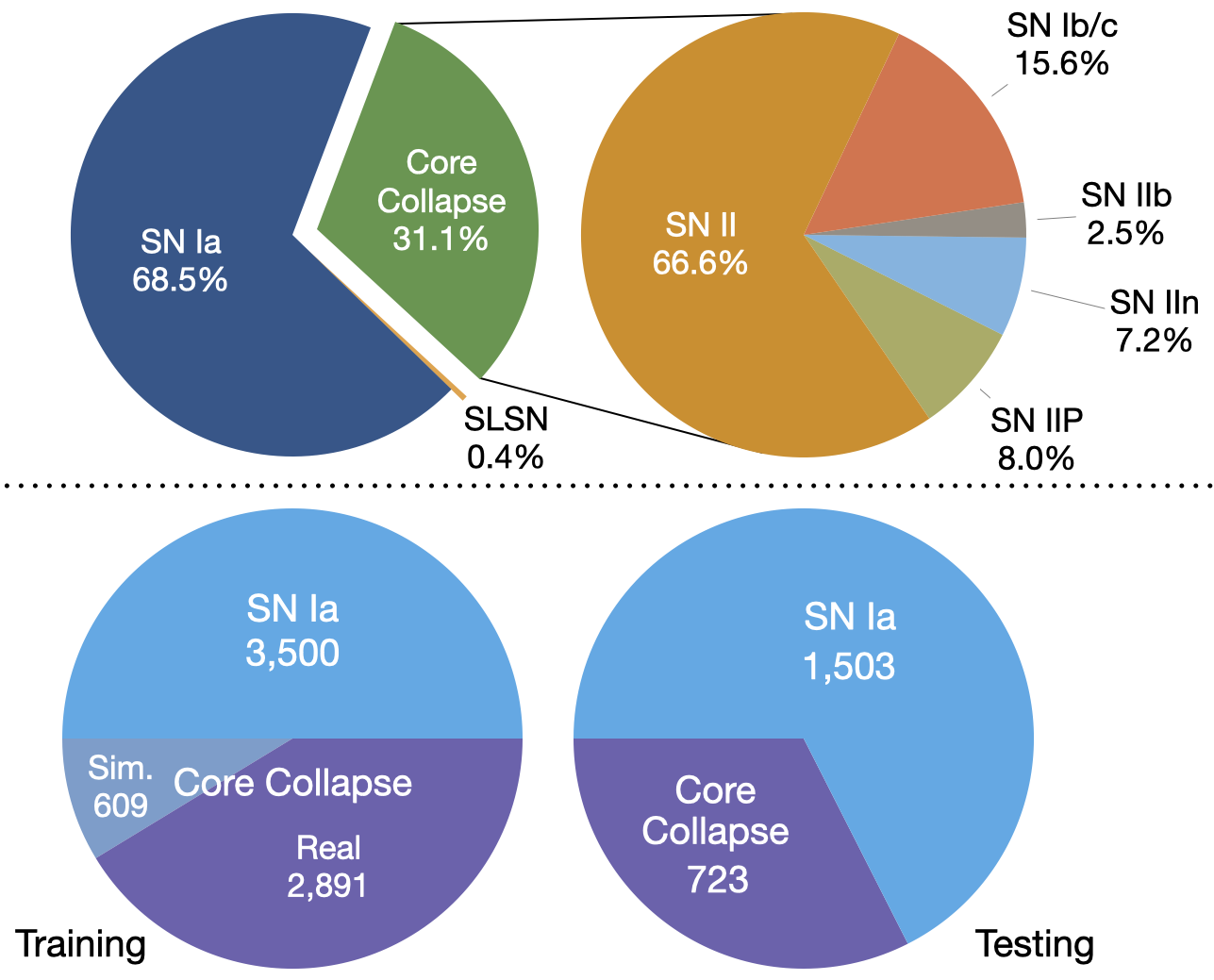}
     \caption{\textbf{Above.} The fraction of the \texttt{GHOST} database comprised by different classes of supernovae. We show the fraction of core-collapse supernovae comprised by different sub-types at right. \textbf{Below.} The total number of supernovae within the training and testing samples for each fold of our two-class model. We have under-sampled our largest class (SN Ia) and over-sampled our smallest class (core-collapse) to generate a balanced dataset for training our algorithm. We list the number of real and augmented core-collapse events for clarity. Due to the small number of peculiar core-collapse, peculiar SN Ia, and SLSN events discovered to date, we are unable to robustly classify these groups by host galaxy properties.}
     \label{fig:PieCharts}
 \end{figure}

We first use k-fold cross-validation to generate five samples of equal size. One of these samples is chosen as the test data in our first random forest model, and the remaining 80\% of the data is used as our training set. To re-balance our training data, we undersample our largest class (SN Ia) and oversample our smallest class (core-collapse) to a size between these two extremes. This allows us to train on as much data from the majority class as possible, while minimizing the amount of artificial data we generate for the minority class. After re-sampling, our training data contains 3,500 SN Ia events and 3,500 core-collapse events in a single fold. Our test set contains approximately 1,503 SN Ia events and 723 core-collapse events in each fold (because our full test set is not divisible by five, each fold will contain a slightly different number of events). The next model uses a different one of the five sub-samples for testing and the remaining re-balanced 80\% of the data for training. This process is repeated for each of the five folds, generating five distinct random forests for our binary classification problem. We plot pie charts comparing the classes of our full dataset and our re-balanced samples in Figure \ref{fig:PieCharts}.

We use the \texttt{RandomizedSearchCV} algorithm to determine the optimal hyperparameters for our random forest. Our final model consists of 1,400 trees without bootstrapping, and considers a maximum of 18 features (the square root of the total number of features) when choosing the best split for each node. We use a maximum depth of 90 and a minimum of 2 samples required for splitting. As with the dimensionality reduction described in \textsection \ref{DimensionReduction}, we scale our features to have a mean of $0$ and variance of unity before classification.

\subsection{Random Forest: Results}\label{Results}
\begin{figure*}[!ht]
   % \captionsetup[subfigure]{labelformat=empty}
     \centering
      \subfloat[][]{\includegraphics[width=0.45\linewidth]{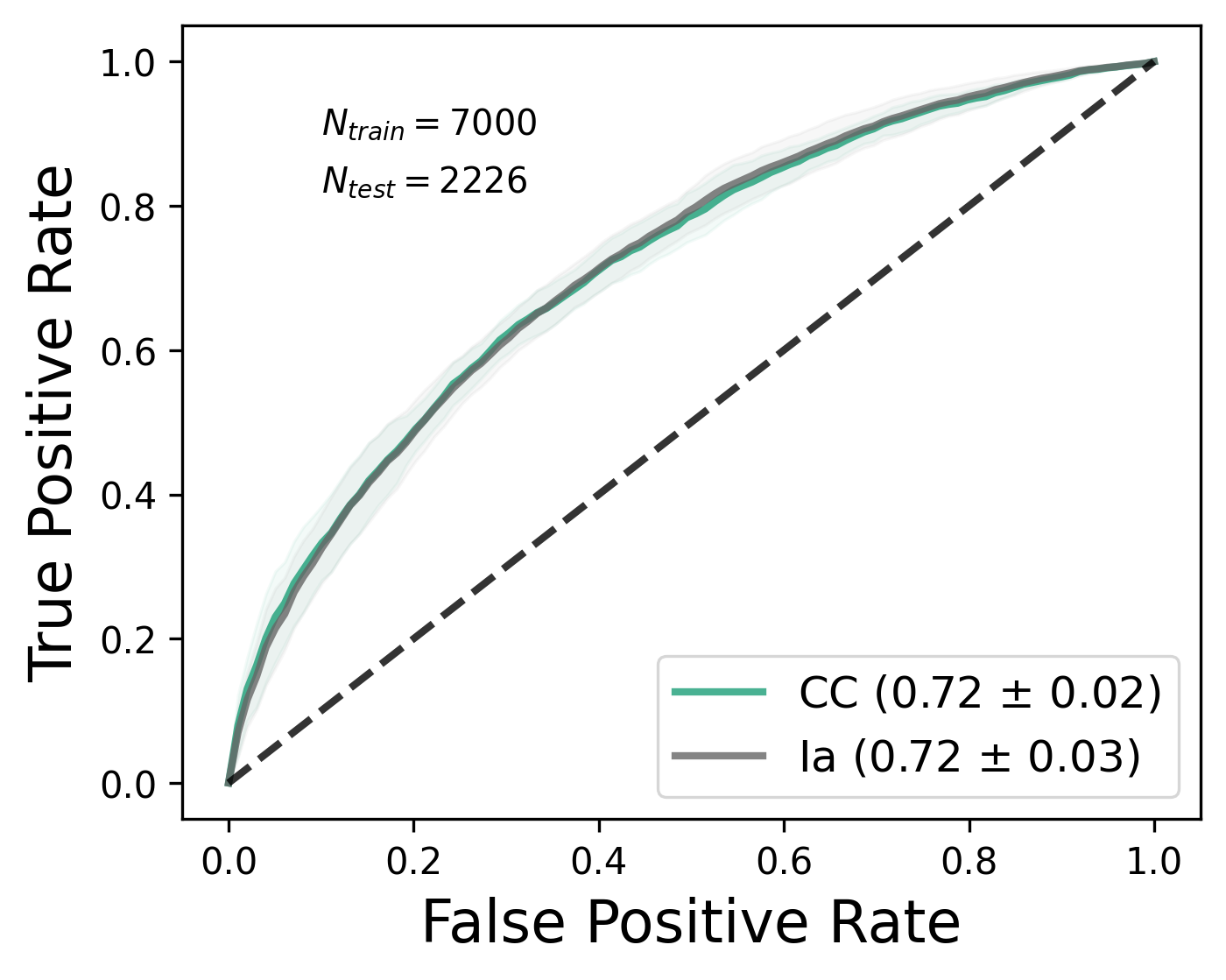}\label{fig:2Class_ROC}}
     \subfloat[][]{\includegraphics[width=0.5\linewidth]{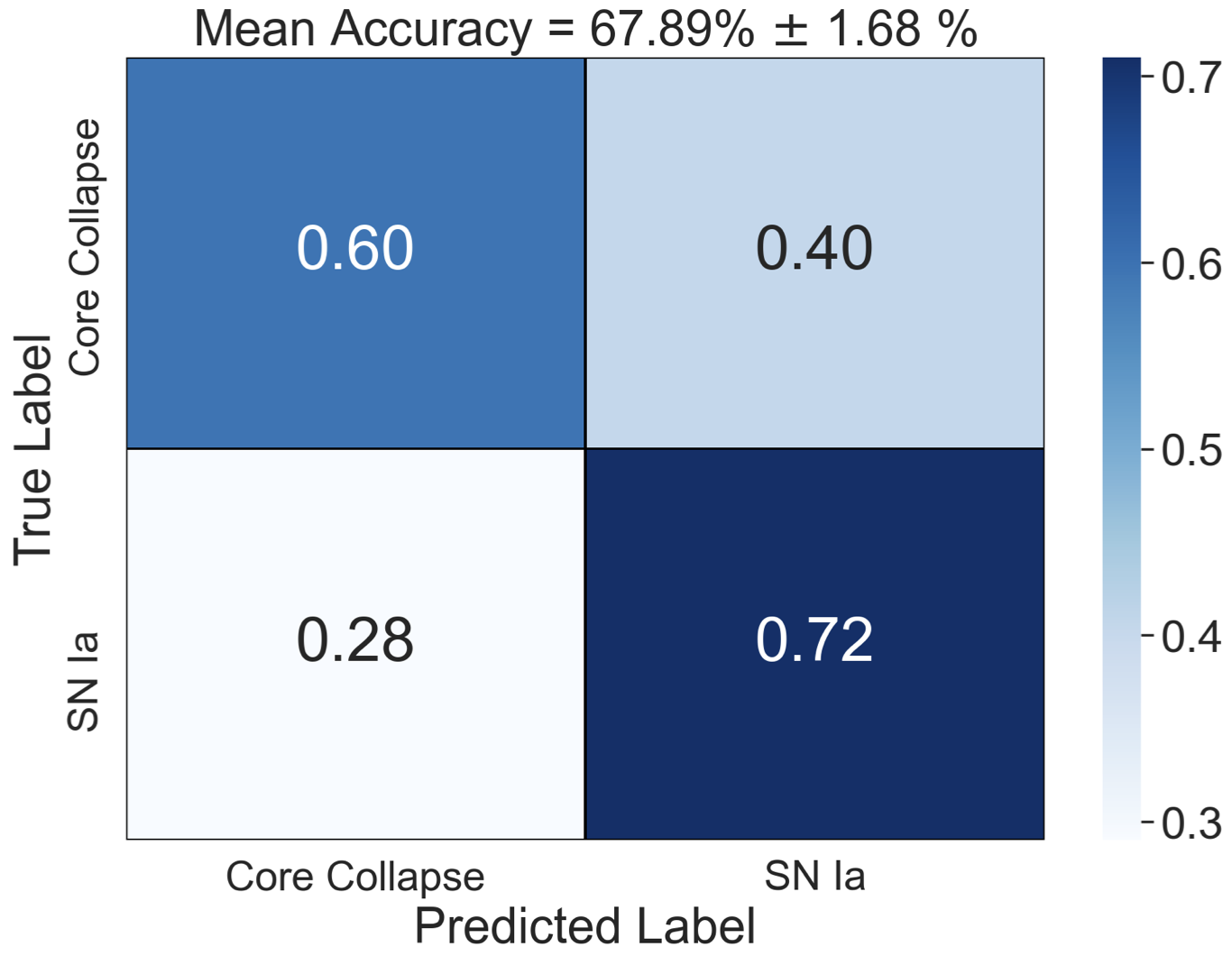}\label{fig:ConfusionMatrix}}
     %\vspace*{-5mm}
     \caption{\textbf{a.} A receiver operator characteristic (ROC) curve for our random forest classifier trained to classify SNe Ia and core-collapse events. Shaded regions describe the sample standard deviation calculated from a 5-fold cross-validation. The dashed line designates the performance of a model that classifies at random, and the area under the curve (AUC) for each class is listed along with its standard deviation across the 5 folds. The total number of supernovae in the training and testing sets (after augmentation) for each fold is listed top left. \textbf{b.} The confusion matrix for our classifier, showing the mean true positive rate and mean false positive rate for each class. The overall accuracy is given at top. We accurately distinguish $\sim70\%$  of SNe Ia and core-collapse events without any information from either the light curve or spectrum.}
     \label{fig:ConfusionMatrix_andROC}
\end{figure*}

Because an event's predicted class is determined by consolidating the predictions of each of the trees, the fraction of final votes belonging to each class can be used as a classification probability. Traditionally, the overall accuracy in a binary classification problem is determined by the fraction of events correctly classified by a majority of trees, corresponding to a decision threshold of $50\%$; however, different thresholds can be selected to prioritize particular aspects of the model.  For example, a higher decision threshold will consider only the events which are classified nearly unanimously by all of the trees, as was the case for our star-galaxy classifier, whereas a low decision threshold will do the opposite. This tradeoff can be conceptualized with a Receiver Operating Characteristic (ROC) curve, which describes the rate of true positive classifications and false positive classifications as a function of the decision threshold (from 0 to 1). The accuracy of our model, which is determined by a majority vote, represents a single point along this curve. The Area Under the ROC Curve (AUC) quantifies the separability of our two classes, with a high AUC indicating a model that achieves a high true positive rate and a low false positive rate. In the limit of perfect classification, the AUC approaches unity. We have constructed a ROC curve in Figure \ref{fig:2Class_ROC} to evaluate the binary model described above. We also report the AUC for each class and its standard deviation across our 5-folds. We present the confusion matrix for our model in Figure \ref{fig:ConfusionMatrix}, which lists the mean classification accuracy (corresponding to a probability threshold of $50\%$) for each class and for the complete test set. 
 
\textit{We find that we can predict supernova class with $\sim 70\%$ accuracy without any spectroscopic or photometric information from the explosion itself}, using exclusively host galaxy information and the angular offset of the supernova. The model achieves an overall accuracy of 67.9\% $\pm$ 1.7\%, with a classification accuracy of 72\% for SN Ia events. The AUC for our final model is 72\% $\pm$ 2\% and 72\% $\pm$ 3\% for SN Ia and core-collapse events, respectively. We also achieve a SN Ia model precision of 79\%. Imputing missing values using the default mean imputation configuration of \texttt{SimpleImputer} from the \texttt{sklearn} package results in a comparable mean accuracy and a larger standard deviation of $66.8\pm 4.6$ \%. 

\begin{figure}
    \centering
    \includegraphics[width=\linewidth]{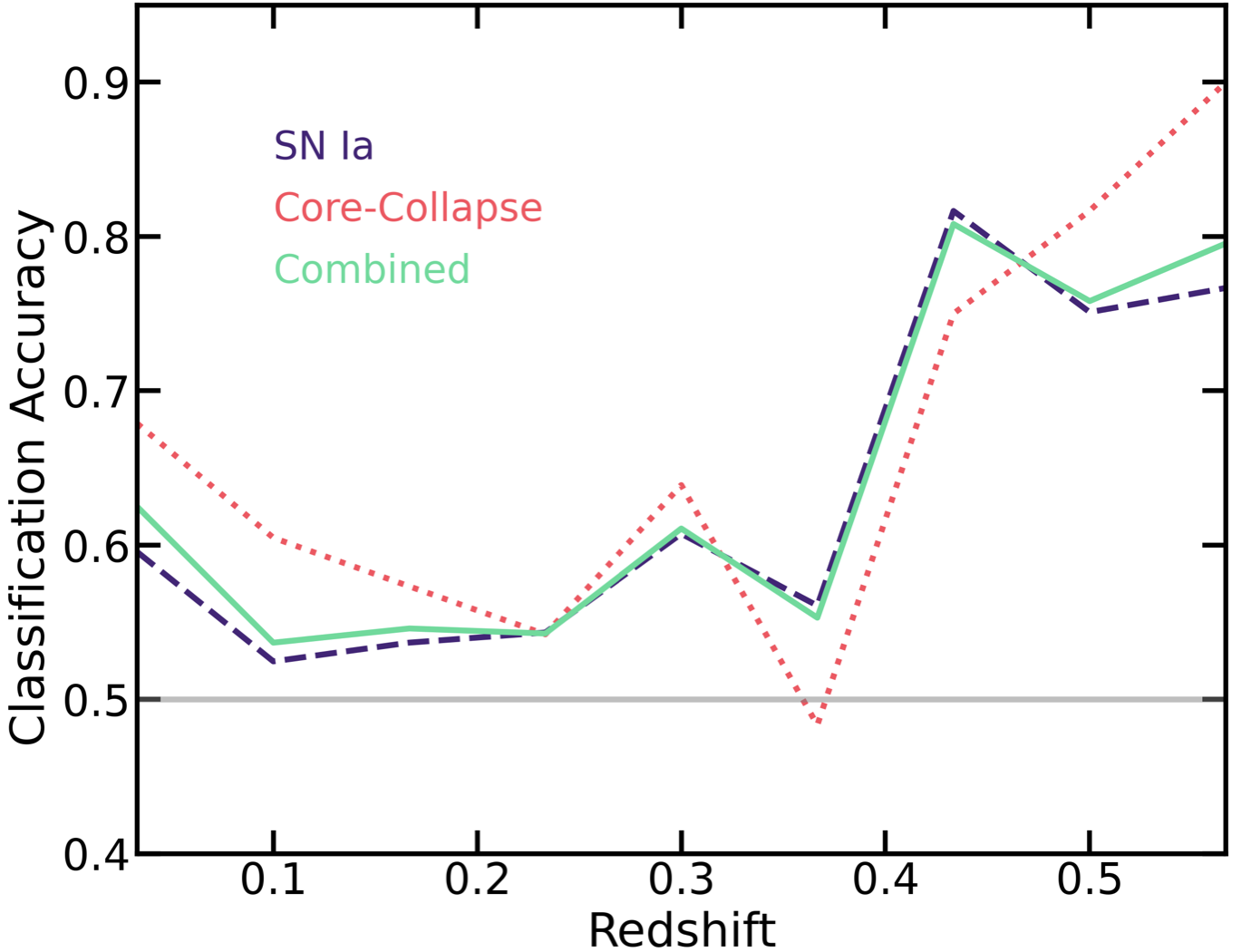}
    \caption{The SN Ia (dashed purple line), core-collapse (dotted red line), and combined classification accuracy (solid green line) of our random forest model as a function of redshift. Supernovae are grouped into 9 bins spanning the range $0 < z < 0.6$. The mean classification accuracy increases nearly monotonically for $z >0.1$; whereas the lowest-redshift events are classified with $\sim60$\% accuracy, $\sim80\%$ of the highest-redshift events are accurately classified. The mean accuracy of the model is higher than random guessing (the solid gray line) for every redshift bin; core-collapse classification does better than random guessing for all but one redshift bin.}
    \label{fig:Acc_vs_z}
\end{figure}

To assess the predictive power of host galaxies as a function of redshift, we divide our events into 9 equally spaced bins spanning $0 < z < 0.6$. We divide the data for each bin into 5 folds, re-balance our training set, and train and test our models. Because of the small number of observed core-collapse events at high redshift (fewer than 100 above $z=0.3$), we re-balance in each bin by only undersampling our SNe Ia events. This prevents our training set from being dominated by simulated data. We plot the classification accuracy at each redshift for SNe Ia, core-collapse events, and across all events in Figure \ref{fig:Acc_vs_z}.

Our mean classification accuracy increases roughly monotonically with redshift after $z > 0.1$, and traces the classification accuracy for SNe Ia because they comprise the majority of events in each bin. In addition, our accuracy for each class in the highest redshift sample is nearly $20\%$ higher than our classification accuracy for these classes at the lowest redshift sample. At every redshift, we achieve a higher mean classification accuracy than random guessing. Despite the small number of events within the highest redshift bin (107 branch-normal SNe Ia and 18 core-collapse events for $0.53 < z < 0.60$), we accurately identify 90\% of core-collapse events and 76.7\% of SNe Ia. By removing the events classified as core-collapse supernovae across all folds, we increase the SN Ia fraction of our high-redshift ($0.53 < z < 0.60$) sample from 85.6\% to 92.2\%. This $\sim7\%$ increase suggests that host galaxy information can be used to increase the purity of SN Ia samples at high-$z$, but because we are unable to associate the majority of supernovae past $z\sim0.6$ the accuracy of reported classifications will likely decrease for events more distant than this.

For comparison, we develop two ``wishful thinking" classifiers that randomly guess the supernova type of each event. The first classifier guesses each class 50$\%$ of the time, and the second guesses ``SN Ia" 100\% of the time. When applied to our test sample, these classifiers achieve mean accuracies of 50\% and 68\% respectively. While the latter has a comparable mean accuracy to our random forest classifier, it achieves a model precision of 68\% for SNe Ia. This makes our random forest classifier $\sim20\%$ more accurate than chance and 11\% more precise at classifying SNe Ia than a model trained on SN Ia event rates. This study is the first to accurately distinguish thousands of SNe Ia and core-collapse supernovae with the photometric properties of their host galaxies. We provide a single random forest classifier, trained on the full \texttt{GHOST} database, as a submodule within the \texttt{astro\_ghost} package. 

We have found across several iterations that our models trained on data spanning the full redshift range of the database classify SNe Ia more accurately than core-collapse events. This trend is reversed in our redshift-binned models: we classify core-collapse supernovae more accurately than SNe Ia for the majority of our redshift samples. Although is possible that our simulated core-collapse events within the full sample were unphysical, it is more likely that the full sample captures more of the diversity of core-collapse events and their host environments \citep{kelly2012core}.  This is supported by the wide spread of SN II in Fig. \ref{fig:joyplot}, although a larger sample of events is needed to probe these host galaxies in more detail. 

The classifier's improved accuracy with redshift is surprising, as more host galaxy information is available at low-$z$.  In comparing the properties of SN II and SN Ia host galaxies in our highest redshift bin, we find that the high-$z$ SNe Ia extend to significantly higher $\theta/d_{DLR}$ values than the high-$z$ SNe II. This parameter is the most significant feature in our high-redshift model, suggesting it is responsible for the model's improved accuracy. If more supernovae were incorrectly matched to neighboring bright galaxies at high-$z$ because their true host galaxies are faint, they will be overwhelmingly SNe Ia and the reported scaled offsets of this group will be shifted upward. We nevertheless find that this effect remains after considering only supernovae that match their host galaxy's redshift to within 5\%. 

Our database has been consolidated from multiple supernova surveys. Supernovae discovered from SDSS-II are included in this analysis, which span the range $0.05<z<0.4$. Past $z\sim0.4$, our samples consist primarily of transients from deeper surveys such as ESSENCE \citep{2007Miknaitis_ESSENCE} and SNLS \citep{2006SNLS}, which extend past $z\sim0.7$. These high-$z$, large-aperture telescope surveys allow us to probe the innermost regions of supernova host galaxies, where faint core-collapse supernovae would typically be masked by galactic extinction. In addition, because many of these high-$z$ surveys are untargeted, we expect to find more SNe Ia in the extended halos of their host galaxies where older stellar populations are found. This effect would also shift the scaled angular offsets of SNe Ia upward. We find by plotting $\theta/d_{DLR}$ as a function of redshift that the median scaled angular offsets of both core-collapse and type Ia supernovae in our sample are roughly comparable for $0.0 < z < 0.35$, and past this point the scaled offsets for SNe II shift to systematically lower values. Further, the scaled offsets for SNe Ia rapidly increase following $z\sim0.4$ and then gradually return to a value comparable to the low-$z$ sample as we begin to reach the observing limits of the high-$z$ surveys. These trends explain both the rapid increase in classification accuracy at $z\sim0.4$ seen in Fig. \ref{fig:Acc_vs_z} and the overall increase in accuracy with redshift, as more of the low-offset core-collapse events are considered in our analysis. Nevertheless, with only 18 core-collapse events in our highest-redshift bin, more events are needed to validate and further explore these effects at higher redshifts.

As in \textsection \ref{tSNEResults}, we can compare the redshift-corrected photometry of these classes to distinguish observational and intrinsic differences between host galaxies. Re-running our classifier using SDSS absolute magnitudes, we find a mean accuracy of 70.5 $\pm$ 1.5\% but a lower core-collapse accuracy of 53.4 $\pm$ 8\%. This result suggests that redshift information plays a larger role in classifying core-collapse events than SNe Ia.

The accuracy of our classification results suggests that host features can be used to minimize contamination within photometric SNe Ia samples. Because these photometric measurements will be already be made by upcoming surveys, this represents a low-overhead strategy for immediately improving cosmological estimates.
 
We have found in section \ref{PCA} that each of our host features does not provide unique information, so we can reduce the complexity of our classification model while sacrificing minimal accuracy. We achieve this by considering only the 13 features with high loading from Figure \ref{fig:PCA_2Comp} in $grizy$. This reduces our parameter space from 317 to 65 features. We use these features to train a new random forest model, and find an accuracy for each class comparable to that reported for the full model.
 
%In addition to considering only the most significant host galaxy features, we can also reduce our parameter space by classifying with only the first ten principal components from \ref{PCA}, which collectively describe $\sim$80\% of the variation in host features. A random forest model trained on this data achieves mean accuracies of $67$\% and $69$\% for SN Ia and core-collapse supernovae, respectively. The comparable accuracies of these models compared with the model trained on our full dataset emphasizes the importance of the features from \ref{PCA} in characterizing the properties of supernova host galaxies. These components are trivial to calculate within an real-time alert broker such as ANTARES, and can dramatically increase the accuracy of real-time transient classifications. 

Despite the differences in host galaxy distributions found in Fig. \ref{fig:joyplot}, we are unable to construct a random forest that can accurately distinguish between rarer sub-classes of supernovae, even after augmenting these classes as was done for the core-collapse events. These rarer events include SLSNe, SNe IIP, SNe IIb, and SNe Ib/c. It is likely that there are too few of these events for our random forest to identify meaningful relationships with their host galaxies. At present, \texttt{GHOST} contains under 100 SLSNe. LSST in Wide-Fast-Deep mode is predicted to find $\sim10^4$ SLSNe per year \citep{villar2018superluminous}, dramatically expanding this dataset and facilitating future studies into the host galaxies of these underrepresented classes. 
 
 %It is possible that these morphological properties are proxies for age, as hosts at high-redshift reveal little morphological information; however In order to constrain this bias, we have only shown this plot for only hosts in the local universe ($z < 0.1$). 

\begin{figure}
    \centering
    \includegraphics[width=\linewidth]{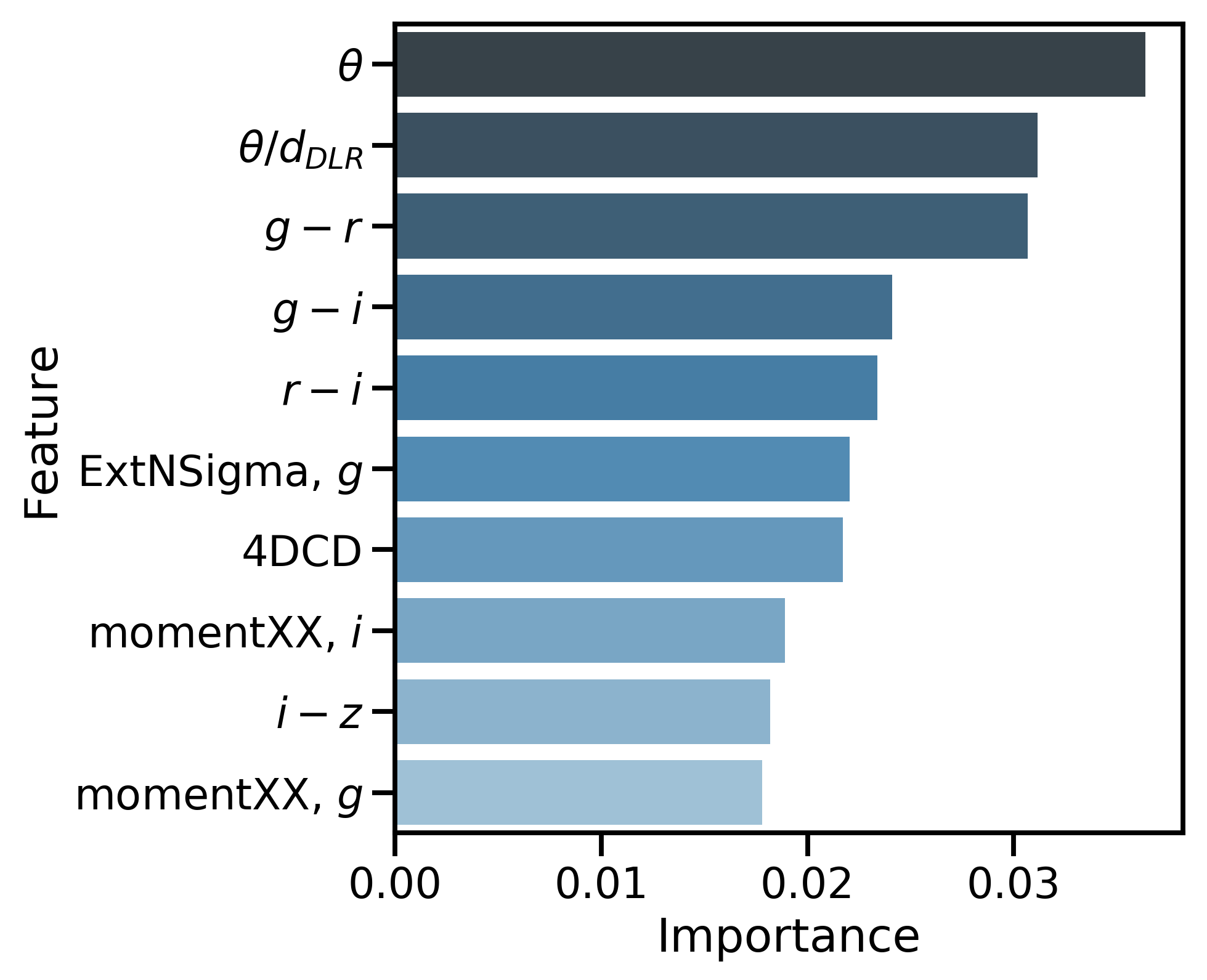}
    \caption{The ten host galaxy features with highest normalized importances in our binary supernova classification model as determined by a gradient boosting model. The features are ordered by importance. The features with highest importance can be broadly categorized by host galaxy color, radial separation, and host galaxy light profile. These features reflect previously-identified host galaxy correlations with supernova type. All of these features are available a-priori or can be calculated immediately following a triggered event.}
    \label{fig:RF_variableBarPlot}
\end{figure}
 
Because of the strong correlations between the majority of host galaxy features, the variable importances returned by our random forest model may not reveal the most valuable features for supernova classification. Further, the random forest importances presented by \texttt{scikit-learn} are known to be biased \footnote{\href{https://explained.ai/rf-importance/\#intro}{https://explained.ai/rf-importance/\#intro}}. We use the package \texttt{feature-selector} to remove all features that are greater than 80\% linearly correlated and train a gradient boosting model for supernova classification. We use the normalized feature importance, defined as the percentage of times the feature is used within the model, as our metric for feature significance. We list the ten most significant features from our model in Fig. \ref{fig:RF_variableBarPlot}.
%with a logloss objective function

We have additionally calculated the permutation importance of our host galaxy features using the \texttt{rfpimp} package. This method identifies the top five features in our model to be $g-r$, second order moments in $g$ and $r$, $ExtNSigma$ in $z$ and $psfMajorFWHM$ in $i$. These results agree with the gradient boosting features and emphasize the importance of color and derived morphological features.

The host galaxy parameters found to be most significant by our gradient boosting model can be grouped into three main categories: radial offset, including $\theta$ and $\theta/d_{DLR}$; color-derived features, including $4DCD$, $g-r$, $g-i$, $r-i$, and $i-z$; and morphological features, including $momentXX$ in $g$ and $i$ and $ExtNSigma$ in $g$. The high importance of scaled radial offset indicates that supernova classes may preferentially occur at different locations throughout their host galaxies. The unscaled radial offset also has high importance, but this is likely a consequence of the difference in observed rates of supernova classes at different redshifts. For example, core-collapse events within \texttt{GHOST} are found predominantly at low redshift, where their observed angular offsets span a wide range of values due to the decreased distance to the host galaxy.

The morphological features characterize the light profile of the host galaxy, which can reflect the classification of the host galaxy along the Hubble tuning fork \citep{van2008dependence}. Color features encode host galaxy metallicity, mass, and galactic star formation rate, which are also correlated with supernova type (e.g. \citealt{10.1111/j.1365-2966.2012.21659.x}). Color is also a useful indicator of early and late-type galaxies (\citealt{strateva2001color}; \citealt{nair2010catalog}). These results suggest that the host galaxy features that are most valuable in classifying supernovae reflect many of the correlations that have been previously identified in the literature. The large PCA loadings of light profile metrics from \textsection \ref{PCA} further confirm the discriminatory power of host galaxy morphology. Finally, the color with highest importance in our model is $g-r$. This color is strongly correlated with host galaxy magnitude (Fig. \ref{fig:g_Corr}), which also has a strong loading in PCA space.

%This study is the first to distinguish thousands of previously discovered supernovae by host galaxy informatio/n alone.
\subsection{A Comparison to Previous Classification Efforts Using Host Galaxy Information}\label{litcomparison}

A similar study was undertaken by \cite{baldeschi2020star} to distinguish SNe Ia and core-collapse events from LOSS \citep{2000Li} and ZTF-BTS \citep{2020Fremling_BTS}, and these samples cover $z \leq 0.15$ and $z \leq 0.005$, respectively. The number of total events classified within each sample was $\sim500$, whereas we classify $>10,000$ total supernovae across all five-folds of our classification scheme. Our train and test sets are randomly selected with replacement across folds, so these are not each unique events; however, we have classified at minimum $2,000$ unique supernovae (the number within a single fold). Because our analysis includes significantly more events spanning $z \leq 2$, we have validated the robustness of host-galaxy correlations for supernova classification across a wider range of redshifts and host galaxies than previous studies. \cite{baldeschi2020star} also included peculiar Ia events (e.g. Ia-02cx,  Ia-91T, Ia-91bg, Ia-SC, Ia-CSM) within their Ia sample; the sample purities for branch-normal events are unknown. Peculiar SN Ia will be unusable for cosmological analysis because of the difficulty of standardizing their photometry. The classification accuracies presented in the current study have only considered branch-normal SNe Ia for classification with the motivation of studying dark energy in upcoming synoptic surveys. Finally, \cite{baldeschi2020star} considered only host galaxies with complete Kron information and used $g$-band Kron radius in PS1-DR2 to associate transients with host galaxies. Because we have developed an association method that is fault-tolerant to faulty PS1 Kron radius estimates (described in \textsection \ref{Methods:PS1Issues}), we are able to accurately identify a host galaxy even when Kron radius values are incorrect or incomplete. Nevertheless, their work reflects the value of using derived host galaxy properties to classify supernovae. This paper has considered the problem of classification directly from observed photometry (and morphological moments), but we will explore derived features such as host metallicity and stellar mass in future work. Their released database of Hubble classes and star formation rates for all Pan-STARRS sources is ideal for extending this work.

In constructing a supernovae classifier from host galaxy information, \cite{foley2013classifying} found that the most significant feature for increasing their FoM (which prioritized SN Ia sample purity) was host morphology. This data increased the FoM by greater than a factor of two over using no host galaxy information. This agrees with our finding that the light profile of a host galaxy ($ExtNSigma$ and $m_{\rm Ap, i} - m_{\rm Kron, i}$ adds significant information for supernova classification. After morphology, \cite{foley2013classifying} find that host galaxy color and luminosity slightly increase their FoM and increase the separation between SNe Ia and core-collapse samples. While this is considered less significant than morphology for maintaining pure SN Ia samples, the significance of color and luminosity in distinguishing supernova classes is supported by the results from our random forest model. Finally, \cite{foley2013classifying} found that radial offset information produces only marginal contributions to SN Ia sample purities. It is surprising, then, that our random forest model uses radial offset information as its most important feature set for classification. We hypothesize that this difference can be attributed to our larger number of supernova events, which may show systematic differences in radial offset that may not be apparent in smaller low-redshift samples. In addition, our core-collapse sample consists of a greater number of sub-types (e.g. IIP, IIn) than were originally considered by \cite{foley2013classifying}, and these sub-types may be more distinguishable by radial offset than SNe Ib/c, SNe II, and SNe Ia only.

Other classifiers have accurately distinguished transient events by their postage stamps alone, and these will inevitably include the photometric properties of the host galaxy; however, these have typically been used to distinguish supernovae from non-explosive transients (\citealt{2019Carrasco}; \citealt{2020Gomez}). These works also train neural networks on complete sequences of images taken throughout the duration of a transient event, and as a result their ability to accurately classify events from real-time single-epoch observations remains unclear. Further, the model developed by \cite{2019Carrasco} requires at least three observations in $g$-band for accurate recall. Although this method is well-equipped for ZTF's public survey cadence of $\sim$3 days in $g$ and $r$, these observations will span nearly a month of the wide-fast-deep survey's single-band cadence. This is far too long to be leveraged for early-time follow-up without retraining. These methods will be more valuable for classifying events within archival data without the need to manually extract photometric measurements. They can also be used to validate pre-maximum classifications made by \texttt{GHOST}. In the event that a supernova that was missed by real-time classifiers is later discovered by archival image sequences, \texttt{GHOST} can still be run in tandem with these methods to identify the host galaxy and provide a classification. 

\texttt{GHOST} characterizes the nature of a supernova that has already been discovered, but it cannot be used to determine whether an event has occurred or not. A real/bogus classifier operating on real-time postage stamps would extend the work of \cite{2019Carrasco} and provide a necessary pre-processing step to the \texttt{GHOST} pipeline.

Calibrated aperture flux, $psfFlux$, adaptive source intensity second moments, and their associated uncertainties will be released for each new source detected by LSST within 24 hours of detection. Difference image alerts will also provide the radial offset of a transient from a likely host galaxy. From these data, color, morphology, and radial features can be derived and used for transient classification. From our importances above, these measures are critical for classifying transient events with host galaxy information. In addition, LSST will accompany source alerts with flags to characterize the ``extendedness" and ``spuriousness" of detected objects, and these can improve the current models for star-galaxy separation and host galaxy association, respectively. Because of the LSST baseline cadence in a single passband, these features will only be available in $ugrizy$ after the first $\sim6$ months of operation. During the initial period of LSST commissioning, host photometry can be retrieved from PS1. Our association method runs in $\sim1$ minute, and classification adds a negligible overhead. This is fast enough to operate on the full LSST alert stream so that host and preliminary class information can be provided with a source alert well within LSST's projected 24 hour window. 

\section{Supernova Siblings}\label{siblings}
By determining the host galaxies of the majority of spectroscopically confirmed supernovae, we can compare the properties of supernova siblings: supernovae associated with the same host galaxy. Extensive work has been done to compare the light curves of SN Ia siblings (\citealt{2018Gall, scolnic2020supernova}) and conduct a census of supernova siblings (\citealt{1990Guthrie, 2013AndersonMultiplicity}). Our pipeline identifies 304 galaxies in our sample that host 2 supernovae, 37 galaxies that host 3 supernovae, 5 that host 4 supernovae, 4 that host 5 supernovae, and 3 that host 6 supernovae. These are nearly half of the supernova siblings identified by \url{http://www.rochesterastronomy.org/snimages/sndupe.html}. The majority of missing pairs were dropped from our pipeline because a supernova was located greater than $30 \arcsec$ from its host galaxy center; however, increasing our cone search would have increased the number of artifacts in our sample and decreased the accuracy of our matches.

For our identified siblings, we use radial offset $\theta$ to explore whether SNe II and SNe Ia probe distinct regions of their host galaxy. For all galaxies hosting multiple supernovae, we take the difference between the radial offsets of two supernova siblings, $\delta \theta$. Next, we generate a probability density function for $\delta \theta$. We compare $\delta \theta$ distributions for two samples: siblings that belong to the same class, where both are SNe II or SNe Ia (the matched sample); and pairs where one sibling is a SN Ia and the other is a SN II (the nonmatched sample). The sizes of these two samples are roughly equal. Because we are comparing the \emph{difference} in $\theta$ between matched and nonmatched siblings, we minimize the influence of different galaxy sizes, distances, and redshifts. A Malmquist bias is present within this sample, as supernovae within nearby host galaxies will be more easily detected than distant supernovae, especially for SNe II. These host galaxies allow for a greater range of possible $\delta \theta$ values than distant galaxies, so a SN II/SN II match will likely have a larger relative separation than a SN Ia/SN Ia match. This effect is unlikely to mask a systematic difference between $\delta \theta$ for matched and unmatched siblings.

We perform an Anderson-Darling (AD) test to compare the distributions of $\delta \theta$ for matched and unmatched supernovae, and find that at $>99.8\%$ confidence we can reject the null hypothesis that these two samples are drawn from the same distribution ($p = 0.002)$. A Kolmogorov-Smirnov (KS) test, which weights the distribution tails less heavily than the AD test, also finds a significant difference between these distributions at the $99\%$ level ($p = 0.01$). 

\cite{wang1997supernovae} makes a similar comparison between the radial offsets (in kpc) of 197 SNe II and 246 SNe Ia using a KS test, rejecting the null hypothesis that these two classes have the same radial distributions at p = 10\% (or at the 90\% confidence level). \cite{wang1997supernovae} further suggests that the dominant differences between the radial offsets of these two classes arise within the inner 6 kpc of a galaxy. To compare the inner radial distributions of SNe Ia and SNe II between galaxies, we can no longer use $\delta \theta$; instead, we compare the scaled radial offset values $\theta/d_{DLR}$ for all events within the \texttt{GHOST} database. We find no significant difference in $\theta/d_{DLR}$ between SNe Ia and SNe II, either in the full range of values ($p=0.47$ for KS and $p>0.25$ for AD) or within the inner 10\% of a supernova's host galaxy ($p = 0.57$ for KS and $p > 0.25$ for AD). We have found this result to be consistent for the low-$z$ sample ($z<0.014$). We caution that we have calculated the DLR for these galaxies using the Kron radius reported for these objects; If this radius is not representative of the true radius of the galaxy, it would decrease the robustness of this comparison. 

 %\begin{figure}
 %   \centering \hspace*{-5mm}
 %   \includegraphics[width=\linewidth]{figures/scaledDLR_twoSided_forPaper.pdf}%
%    \caption{The difference between scaled directional light radius for matched and unmatched supernova siblings. Our results suggest that type Ia and type II supernovae do not occur at statistically significantly different galactocentric distances at any redshift.}
%    \label{fig:GalaxyPairs_KDEs}
%\end{figure}

\section{Discussion}\label{Discussion}
We have constructed the largest catalog of supernovae and their host galaxies to date. By augmenting the DLR method with a gradient ascent algorithm that isolates galaxy profiles directly from images, we can associate host galaxies spanning a wide range of redshifts ($0.00015 \leq z \leq 2$) to an accuracy of $>97\%$.  LSST will have a maximum all-band wide-field-deep depth of $5\sigma > 25.08$, and the Nancy Grace Roman Space Telescope in deep survey mode is slated to achieve a depth per exposure of $26.2$ in $J$ and $H$ \citep{2018HounsellWFIRST}. These programs will allow us to study supernova host galaxies in unprecedented detail and further explore the correlations between host galaxy and transient, but it will also exacerbate the host galaxy association issues described in \textsection \ref{Methods:PS1Issues}. A light profile-based association method will allow us to leverage the strengths of these surveys to continue to accurately associate new transients.

Our results indicate that host galaxy color and photometry can aid in supernova classification. By projecting our sample into reduced parameter spaces, we can immediately identify visual differences between supernova host galaxies. Using a random forest classifier trained on host galaxy information, we distinguish SNe Ia and core-collapse supernovae with $\sim70\%$ accuracy. This work builds on previous efforts by incorporating a significantly larger number of host galaxy features than were considered by \cite{foley2013classifying}.

Our mean classification accuracy is $\sim20\%$ higher for high-$z$ events than for low-$z$ events. This likely reflects differences in the survey strategies used to discover supernovae in these two regimes, but at present we do not have enough high-redshift events to statistically  compare these two samples. Our publically available classifier is trained on the full dataset, which has a median redshift of $0.07$. We predict comparable classification accuracies to those listed above for $z < 0.6$ events. Re-training our algorithm with additional high-$z$ host galaxies, whether simulated or discovered in the first few months of LSST, will ensure that our classifier is robust for more distant events.

The host galaxy features with the most discriminatory power are radial offset, color, and derived morphological features. The color and morphology of a galaxy are indicative of its metallicity and star formation rate, both of which can be used to distinguish SNe Ia and core-collapse supernovae. Accurate morphology estimates in large surveys will be crucial for taking advantage of these correlations for new events. Further, the importance of radial offset suggests that a transient's local environment may be even more predictive of supernova class than the global properties of its host galaxy. The strong overlap between global features of SN Ia, SLSN, and SN Ib/c host galaxies in Fig. \ref{fig:PCA_2Comp} reinforces this need for localized metrics to characterize the local environments of supernovae and more accurately distinguish SN classes.

%which will perform initial alert prioritization and photometric classification on the full LSST alert stream.
This study will inform the development of brokers for the LSST Dark Energy Science Collaboration (DESC). Brokers will need to perform low-latency, accurate classification of transient events using the information provided by the real-time alerts stream. For LSST, these alert packets will include the supernova radial offset, a photometric redshift, and a postage stamp of the field. After using this information to verify the host galaxy, the broker can retrieve archival LSST (or PS1, if it is not available) photometry for that host galaxy, calculate the color of the host, and use the random forest model described above to predict the class of the supernova. By augmenting this algorithm with a real-time photometric classifier such as RAPID \citep{muthukrishna2019rapid}, we can ensure accurate classification pre-maximum for a broad range of transient events.

In addition, we find a statistically significant difference between the differential radial offsets $\delta \theta$ of matched (SN Ia, SN Ia and SN II, SN II) and nonmatched (SN Ia, SN II) supernova siblings, but we do not find a significant difference in the scaled radial offsets $\theta/d_{DLR}$ between SNe Ia and SNe II. Intuitively, one might expect these distributions to differ because core-collapse supernovae are associated with star-forming regions and frequently found in the arms of spiral galaxies. Our result suggests that offset alone may not sufficiently characterize the positions of star forming regions. In addition, the significance of radial offset in our classification model suggests that sub-types of core-collapse events (SNe IIP, SNe Ib/c, SNe IIn) may differ from SNe Ia in radial offset even if SNe II do not. This result underscores the need for higher resolution imaging to assemble a more complete picture of host galaxy-supernova interactions at the local level. This result may also reflect the limitations of characterizing host galaxy morphology using its PS1 Kron radius. As we discuss in greater detail in \textsection \ref{Methods:PS1Issues}, this value is unphysical or unreported for the largest and smallest host galaxies in our sample, limiting our ability to directly compare SN positions between these galaxies.

Finally, we find systematic differences in the host galaxy photometry of underrepresented events such as SLSNe, SNe IIP, and SNe IIb. By comparing rest-frame and absolute $g$-band magnitudes, we identify a strong redshift-dependence in the rates of these events. Archival SLSNe are found at higher redshifts than SNe IIb, and as a result their observed host galaxies will be fainter. This information can be used to distinguish these events and validate photometric redshift estimates.

LSST's official strategy for host galaxy association has not yet been finalized. We have shown that, where deep surveys limit the ability to de-blend extended sources, the directional light radius is unable to robustly identify supernova host galaxies. In addition, the postage stamps associated with real-time transient alerts for LSST are unlikely to exceed $6\arcsec \ \textrm{x} \ 6\arcsec$. This window will exclude many supernova host galaxies within z$\sim$0.1 (See Figure \ref{fig:Redshift_mismatch}), and make it difficult to use postage stamp gradients for host galaxy association. The gradient ascent method outlined in this work will be a valuable resource for validating LSST's host galaxy associations and proposing improvements to the pipeline.
%\newpage
\section{Future Work}\label{FutureWork}
The \texttt{GHOST} database can facilitate future studies into the correlations between supernovae and their host galaxies. The LSST DESC Science Roadmap lists a vital collaboration goal as identifying the underlying physics of the SN-host galaxy mass correlation for cosmological analysis. A strong correlation has been identified between local stellar mass and distance residuals for low-$z$ Ias, corrections for which would reduce the uncertainties placed on H$_0$ \citep{jones2018should}; further, a local color step derived from 2MASS photometry of a sample of Type Ia supernovae was found to be significant to 7$\sigma$ \citep{roman2018dependence}. A similarly significant (5.7$\sigma$) correlation was identified as a function of specific star formation rate \citep{rigault2018}. Our database contains $>9,200$ SNe Ia, and by increasing the sample size of associated SNe Ia by at least an order of magnitude (compared to e.g. \citealt{2018SDSS}) this data will shed more light on these correlations.

Our tSNE results suggest there are systematic differences in the host galaxies of supernova sub-types that we have not yet incorporated into a classification model. These redshift-dependent differences may not be significant enough for accurate classification by themselves, but are likely to improve the accuracy of photometric classifiers. We will explore this possibility when developing an ensemble host photometry-SN photometry classifier in the future.

In addition, we will continue improving our gradient ascent algorithm. We have avoided incorporating spectroscopic redshift information into the algorithm so that the method may be automated on the alert streams of large surveys where estimates may have large uncertainties, but it is likely that knowledge of a photo-$z$ would inform both the use of the method over DLR and the selection of a relevant step size. An accurate pipeline for host galaxy association will enable discoveries across a broad range of time-domain studies, such as by constraining kilonova progenitor models \citep{jiang2020simulating}, verifying the discovery of Tidal Disruption Events \citep{french2016tidal}, and further uncovering the relationship between Rapidly Evolving Transients (RETs) and their host galaxies \citep{wiseman2020host}. 

We will next explore the use of Convolutional Neural Networks (CNNs) to classify supernovae using postage stamps of their host galaxies. Postage stamps encode spatially resolved color and brightness information, such as radial color gradients \citep{park2005morphology}, that may be more accurate in classifying transients than the single-statistic PS1 features we have considered here. CNNs have already achieved high accuracies distinguishing simulated SNe Ia from other supernovae (with an AUC of 96\% in a single-epoch; \citealt{kimura2017single}), and the \texttt{GHOST} database is ideal for extending this work. Finding the global and local correlations that can improve classification is also critical for plausibly embedding supernovae in LSST-depth postage stamps. This simulated data will be used to train photometric classifiers in preparation for LSST, another primary goal for LSST DESC. By extracting host galaxy features directly from postage stamps in a neural network, we will be able to more accurately encode these correlations in our simulated sample.
%\newpage
\section{Acknowledgements}\label{acknowledgements}
Author contributions are listed below.\\
A. Gagliano: Software, website, and database development; writing and editing.\\
G. Narayan: Oversight, writing and editing. \\
A. Engel: Contributed software for the SN Ia and core-collapse classifier.\\
M. Carrasco-Kind: Contributed software for the \texttt{GHOST} Viewer.\\

This paper has undergone internal review in the LSST Dark Energy Science Collaboration. The internal reviewers were Ryan Foley, Renee Hlozek, Kaisey Mandel, Christian Setzer, and Seth Digel. The authors thank the internal reviewers and the anonymous referee for their thorough reviews of this work. Each has significantly strengthened this paper.  

The DESC acknowledges ongoing support from the Institut National de Physique Nucl\'eaire et de Physique des Particules in France; the Science \& Technology Facilities Council in the United Kingdom; and the Department of Energy, the National Science Foundation, and the LSST Corporation in the United States.  DESC uses resources of the IN2P3 Computing Center (CC-IN2P3--Lyon/Villeurbanne - France) funded by the Centre National de la Recherche Scientifique; the National Energy Research Scientific Computing Center, a DOE Office of Science User Facility supported by the Office of Science of the U.S.\ Department of Energy under Contract No.\ DE-AC02-05CH11231; STFC DiRAC HPC Facilities, funded by UK BIS National E-infrastructure capital grants; and the UK particle physics grid, supported by the GridPP Collaboration.  This work was performed in part under DOE Contract DE-AC02-76SF00515.

AG is supported by the Illinois Distinguished Fellowship, the National Science Foundation Graduate Research Fellowship Program under Grant No. DGE – 1746047, and the Center for Astrophysical Surveys Graduate Fellowship at the University of Illinois. GN's work on this project was partially supported by the Lasker Data Science Fellowship at STScI, and generous startup funding from the University of Illinois. The authors thank the Space Telescope Science Institute for providing travel funds for AG. We are also grateful for the support of the National Center for Supercomputing Applications (NCSA), which houses the server that runs the \verb|GHOST| database. The authors thank Rahul Biswas, Qinan Wang and Monika Soraisam for fruitful conversations related to this work. 

The Pan-STARRS1 Surveys (PS1) and the PS1 public science archive have been made possible through contributions by the Institute for Astronomy, the University of Hawaii, the Pan-STARRS Project Office, the Max-Planck Society and its participating institutes, the Max Planck Institute for Astronomy, Heidelberg and the Max Planck Institute for Extraterrestrial Physics, Garching, The Johns Hopkins University, Durham University, the University of Edinburgh, the Queen's University Belfast, the Harvard-Smithsonian Center for Astrophysics, the Las Cumbres Observatory Global Telescope Network Incorporated, the National Central University of Taiwan, the Space Telescope Science Institute, the National Aeronautics and Space Administration under Grant No. NNX08AR22G issued through the Planetary Science Division of the NASA Science Mission Directorate, the National Science Foundation Grant No. AST-1238877, the University of Maryland, Eotvos Lorand University (ELTE), the Los Alamos National Laboratory, and the Gordon and Betty Moore Foundation. 

This research has made use of the NASA/IPAC Extragalactic Database (NED),
which is operated by the Jet Propulsion Laboratory, California Institute of Technology,
under contract with the National Aeronautics and Space Administration. Funding for the Sloan Digital Sky Survey IV has been provided by the Alfred P. Sloan Foundation, the U.S. Department of Energy Office of Science, and the Participating Institutions. SDSS-IV acknowledges support and resources from the Center for High-Performance Computing at
the University of Utah. The SDSS web site is www.sdss.org. SDSS-IV is managed by the Astrophysical Research Consortium for the 
Participating Institutions of the SDSS Collaboration including the 
Brazilian Participation Group, the Carnegie Institution for Science, 
Carnegie Mellon University, the Chilean Participation Group, the French Participation Group, Harvard-Smithsonian Center for Astrophysics, 
Instituto de Astrof\'isica de Canarias, The Johns Hopkins University, Kavli Institute for the Physics and Mathematics of the Universe (IPMU) / 
University of Tokyo, the Korean Participation Group, Lawrence Berkeley National Laboratory, 
Leibniz Institut f\"ur Astrophysik Potsdam (AIP),  
Max-Planck-Institut f\"ur Astronomie (MPIA Heidelberg), 
Max-Planck-Institut f\"ur Astrophysik (MPA Garching), 
Max-Planck-Institut f\"ur Extraterrestrische Physik (MPE), 
National Astronomical Observatories of China, New Mexico State University, 
New York University, University of Notre Dame, 
Observat\'ario Nacional / MCTI, The Ohio State University, 
Pennsylvania State University, Shanghai Astronomical Observatory, 
United Kingdom Participation Group,
Universidad Nacional Aut\'onoma de M\'exico, University of Arizona, 
University of Colorado Boulder, University of Oxford, University of Portsmouth, 
University of Utah, University of Virginia, University of Washington, University of Wisconsin, Vanderbilt University, and Yale University.

This research has made use of the following \texttt{Python} software packages:
\texttt{Astropy} \citep{astropy:2018}, \texttt{Matplotlib} \citep{hunter2007matplotlib}, \texttt{Pandas} \citep{mckinney2010data}, \texttt{NumPy} \citep{walt2011numpy}, \texttt{Seaborn} \citep{michael_waskom_2014_12710}, \texttt{SciPy} \citep{jones2001scipy}, \texttt{Scikit-Learn} \citep{pedregosa2011scikit}, and \texttt{Imbalanced-learn} \citep{JMLR:v18:16-365}. 

\bibliography{references}

\end{document}